\documentclass[aps,prd,twocolumn,groupedaddress,nofootinbib]{revtex4-1}

\usepackage{graphicx,color,amsmath,amssymb}

\newcommand{\metre}{{\, {\rm m}}}
\newcommand{\cm}{{\, {\rm cm}}}

\newcommand{\eV}{{\, {\rm eV}}}
\newcommand{\keV}{{\, {\rm keV}}}

\newcommand{\GeV}{{\, {\rm GeV}}}

\newcommand{\kelvin}{{\, {\rm K}}}

\newcommand{\MHz}{{\, {\rm MHz}}}
\newcommand{\GHz}{{\, {\rm GHz}}}

\newcommand{\LL}{{\mathcal L}}
\newcommand{\OO}{{\mathcal O}}

\newcommand{\FF}{{\mathcal F}}
\newcommand{\SC}{{\bar S}}

\newcommand{\PP}{{\mathbb{P}}}

\newcommand{\Ja}{{J^{(a)}}}

\newcommand{\tchi}{{\tilde{\chi}}}

\newcommand{\PPex}{{\PP_{\rm ex}}}

\newcommand{\Tesla}{{\, {\rm T}}}

\newcommand{\intinf}{{\int_{-\infty}^{\infty}}}
\newcommand{\intoinf}{{\int_{0}^{\infty}}}

\definecolor{mypurple}{RGB}{164,64,214}

\begin{document}

\title{Parametrics of electromagnetic searches for axion dark matter}

\author{Robert Lasenby}
\email{rlasenby@stanford.edu}
\affiliation{Stanford Institute for Theoretical Physics, Stanford University, Stanford, CA 94305, USA}

\date{\today}

\begin{abstract}
 Light axion-like particles occur in many theories of
 beyond-Standard-Model physics, and may make up some or all
 of the universe's dark matter. One of the ways they can
 couple to the Standard Model is through the electromagnetic
 $F_{\mu\nu} \tilde F^{\mu\nu}$ portal, and there is a
 broad experimental program, covering many decades in
 mass range, aiming to search for axion dark matter via
 this coupling. In this paper, we derive limits on the
 absorbed power, and coupling sensitivity, for a broad
 class of such searches. We find that standard techniques,
 such as resonant cavities and dielectric haloscopes, can
 achieve $\OO(1)$-optimal axion-mass-averaged signal powers, for given volume
 and magnetic field. For low-mass (frequency $\ll \GHz$)
 axions, experiments using static background magnetic fields
 generally have suppressed sensitivity --- we discuss the
 physics of this limitation, and propose experimental
 methods to avoid it, such as microwave up-conversion
experiments. We also comment on the detection of
 other forms of dark matter, including dark photons, as well
 as the detection of relativistic hidden sector particles.
\end{abstract}

\maketitle


\section{Introduction}

Axion-like particles, in particular the QCD axion,
are a well-motivated dark matter (DM) candidate. They occur in many 
models of beyond-Standard-Model physics, and can naturally be
light and weakly-coupled, allowing them to be stable
and difficult-to-detect. There are also a number of early-universe
production mechanisms, which can produce them in the correct abundance
to be the DM~\cite{Kawasaki:2014sqa,Ringwald:2015dsf,Co:2017mop,diCortona:2015ldu,Agrawal:2017eqm,Co:2017mop,Co:2018phi,Co:2018mho,Co:2019jts}.

A wide range of existing and proposed experiments aim to detect
axion DM candidates. These span many decades of mass range,
and target a variety of possible couplings to the Standard Model (SM)~\cite{Graham:2015ouw}.
In this paper, we will focus on the $a F_{\mu\nu} \tilde F^{\mu\nu}$
axion-photon-photon coupling, and address
the sensitivity limits on such experiments ---
how small a DM-SM coupling could we possibly detect,
given the dimensions, timescales, sensors etc. available?
We choose the $a F_{\mu\nu} \tilde F^{\mu\nu}$ coupling partly because,
for a generic QCD
axion, this coupling must lie within a fairly narrow (logarithmic) range~\cite{diCortona:2015ldu,DiLuzio:2016sbl}; 
it is also a generic feature of many other axion-like-particle
models~\cite{Jaeckel:2010ni}. In addition, it represents a particularly easily-analysed
example of the kind of sensitivity limits we are interested in.

We derive bounds on the power absorbed
by axion DM experiments, under fairly general
assumptions, in terms 
of the magnetic field energy
maintained inside the experimental volume.
We also derive related limits on the achievable
sensitivity for such experiments, using the tools
of quantum measurement theory.
These analyses are analogous to those developed
in the gravitational wave detection literature 
(see e.g.~\cite{Miao,Pang2019} and references therein).

For low-mass (frequency $\ll \GHz$) axions, we
review why static-background-field experiments
generally have suppressed sensitivity
(compared to their scaling at higher frequencies),
and point out that this suppression can be alleviated
in a number of ways, potentially motivating new experimental
concepts. In particular, we point out that `up-conversion'
experiments, using a magnetic field oscillating at microwave
frequencies, can have parametrically better scaling
for small axion masses, and discuss the experimental details
in a companion paper~\cite{srf}.

Similar sensitivity analyses can be applied directly
to other forms of DM that couple to the SM photon,
e.g.\ dark photon DM with a kinetic mixing.
Analogous ideas can also be applied to other kinds of
DM-SM couplings, and beyond that, to the detection
of general hidden-sector states. We comment on some
of these extensions later in the paper,
and in the conclusions.

\subsection{Summary of results}
\label{secSummary}

Here, we give a brief summary of our main results.
Suppose that dark matter consists of an axion-like particle $a$ 
of mass $m_a$,
with a coupling $\LL \supset -\frac{1}{4} g_{a\gamma\gamma} a
F_{\mu\nu}\tilde F^{\mu\nu}$ to the SM.
For non-relativistic axion DM, this acts
as an `effective current' $J^{(a)} \simeq g_{a \gamma \gamma} \dot a B$, 
where $B$ is the magnetic field.
Suppose that we have some laboratory system 
with which we will try to detect the axion DM.
If we want to search for axions
over a mass range $\Delta m$, then for
$g_{a \gamma \gamma}$ small enough that the target system
is in the linear response regime,
the expected time-averaged absorbed power from the axion effective current,
at the least favourable axion mass, satisfies
(under assumptions that we will discuss)
\begin{equation}
	\bar P \le 
	\pi	\frac{g_{a \gamma \gamma}^2 \rho_a \overline{U}_B}{\Delta m}
	\label{eqpbound1}
\end{equation}
where $\rho_a$ is the energy density of the axion DM,
and $\overline{U}_B$ is the time-averaged
magnetic field energy
in the experimental volume (ignoring magnetic fields
on very small spatial scales).
Here, the expectation value includes integrating over
the unknown phases of the axion signal (otherwise,
there could be $\OO(g_{a\gamma \gamma})$ components
of the absorbed power). This bound applies
to target systems for which the imaginary part of the
response function is non-negative at positive frequencies
--- that is, the system on average absorbs energy from
the axion forcing, rather than emitting.
As we discuss below, $\bar P$ is, in many circumstances,
closely related to the detectability of an axion signal.
The $\bar P \propto 1/\Delta m$ behaviour
corresponds to the power vs bandwidth
tradeoff that is a property of many detection schemes;
covering a broader axion mass range in the same
integration time necessarily leads to lower average
signal power.

In the $\nu_a \gtrsim \GHz$ regime (where
$\nu_a = m_a / (2 \pi)$ is the frequency 
of the axion oscillation),
cavity haloscopes (as initially proposed in~\cite{Sikivie:1983ip}) such as ADMX~\cite{Du:2018uak,Braine:2019fqb}
and HAYSTAC~\cite{Zhong:2018rsr,Droster:2019fur} can attain the bound in equation~\ref{eqpbound1} to $\OO(1)$. At higher frequencies ($\nu_a \gtrsim 10 \GHz$), dielectric haloscope 
proposals \cite{TheMADMAXWorkingGroup:2016hpc,Baryakhtar:2018doz} 
can also achieve this, taking the experimental volume
to be that occupied by the dielectric layers. 

However, for $m_a \lesssim L^{-1}$, where $L$ is
the length scale of the shielded experimental
volume, the EM modes at frequencies $\sim \nu_a$
are naturally in the quasi-static regime. In that
case, a static-background-field experiment has
\begin{equation}
	\bar P \lesssim 
	\pi \frac{g_{a \gamma \gamma}^2 \rho_a \overline{U}_B}{\Delta m}
	(m_a L)^2
	\label{eqpqs1}
\end{equation}
This suppression affects low-frequency ($\nu_a \ll \GHz$) axion DM detection
proposals such as \cite{Sikivie2014}, ABRACADABRA~\cite{Kahn:2016aff}, and DM Radio~\cite{scshort}. As discussed below, the
scaling of the detectability limit is similarly affected,
with the minimum detectable $g_{a \gamma \gamma}$ increased
by $\sim (m_a L)^{-1}$.
Our results agree parametrically with those of
\cite{scshort,sclong}, which analyse 
axion DM detection in the quasi-static
regime, using an inductive pickup and linear
phase-invariant amplification.

Even under the assumptions leading
to equation~\ref{eqpbound1}, the quasi-static suppression
is not inevitable. To alleviate it for
static background magnetic fields, we would need to enhance
the quantum fluctuations of the EM fields
that couple to the axion effective current,
at frequencies $\sim \nu_a$.
One way to achieve this is for the field fluctuations
to `borrow' energy from some other source,
e.g.\ magnetisation energy in a material,
or a circuit component with negative differential
resistance. The practicality of such concepts
requires further investigation.

The quasi-static suppression can also be alleviated
by performing an `up-conversion' experiment, in which
the background magnetic field is oscillating
at a frequency $\gtrsim L^{-1}$.
Up-conversion experiments have been proposed in the optical 
range~\cite{DeRocco:2018jwe,Obata:2018vvr,Liu:2018icu},
but the relatively small amplitude of achievable optical-frequency
fields means that they would have relatively poor
sensitivity.
Larger magnetic fields are attainable at lower frequencies;
in particular, it is routine to obtain
magnetic fields of $\sim 0.1 \Tesla$ at $\sim \GHz$ frequencies
in superconducting (SRF) cavities~\cite{padamsee1998rf,Grassellino:2017bod}.

These field strengths were noted in~\cite{Sikivie:2010fa}, which
proposed a SRF up-conversion experiment.\footnote{Microwave up-conversion experiments
were also proposed in~\cite{Tobar_2020,Goryachev:2018vjt}, but their sensitivity
projections rely on an alternative interpretation of axion electrodynamics
--- see Section~\ref{secUpExp}.} 
However, they mainly considered $\nu_a \sim \GHz$,
for which static field experiments do not encounter 
the quasi-static suppression.
Consequently, the only benefit of an SRF experiment would be
the higher cavity quality factor, which is unlikely
to overcome the disadvantages of smaller background magnetic field,
higher temperature (due to cooling power requirements), and
drive-related noise issues.
However, as we point out here, for $\nu_a \ll \GHz$,
the lack of quasi-static
suppression may make up-conversion more competitive.
We investigate this possibility in 
more detail in a companion paper~\cite{srf} (see also~\cite{Berlin_2019}).

\subsubsection{Detectability}
\label{secSumDet}

In the above paragraphs, we discussed the average
power absorbed from the axion
effective current. It is obvious that, other things being
equal, a higher absorbed power makes it easier
to detect axion DM. However, in comparing
different experiments, other things are often not equal,
and more generally, it is useful to have quantitative
limits on how small a coupling can be detected.

By using quantum measurement techniques~\cite{Clerk_2010},
we could in principle detect extremely small $g_{a \gamma \gamma}$,
for a given $\overline{U}_B$. For example, by preparing
a cavity mode in a large-number Fock state, we could
Bose-enhance the absorption (and emission) of axions~\cite{chou1}.
However, such techniques are often difficult to implement;
in axion experiments, the only similar measurement
demonstrated so far is HAYSTAC's squeezed state receiver,
which is planned to deliver
a factor 2 scan rate improvement in the forthcoming 
Phase II run~\cite{Droster:2019fur}.

To understand different signal detection techniques, we can
consider splitting our laboratory system into
a `target' part, which couples to the axion effective
current, and a separate `readout' part, which couples 
to the target. 
One common technique used in readout schemes
is linear, phase-invariant amplification 
(in particular, it is employed by almost all
existing and proposed axion detection experiments 
at microwave frequencies and below).
For different experimental setups, we can place
limits on the SNR obtained, in analogy to the
absorbed power limit from equation~\ref{eqpbound1}.

If a linear amplifier is
employed in `op-amp' mode~\cite{Clerk_2010},
and is subject to the `Standard Quantum
Limit' (SQL)~\cite{Kampel:2017gze} (see
appendix~\ref{appQmeas}), then the SNR obtained,
averaged across a fractionally
small axion mass range $\Delta m$, is
bounded by 
\begin{equation}
	\overline{{\rm SNR}^2} \lesssim
	C \pi (g^2 \rho_a \overline{U}_B)^2 \frac{t Q_a}{\omega_1^3 \gamma \Delta m} 
	\label{eqsnr2intro}
\end{equation}
where $Q_a \simeq 10^6$ is the fractional bandwidth
of the axion DM signal (see section~\ref{secAxVel}),
$\omega_1 \simeq m_a \pm \omega_B$ is the oscillation
frequency of the axion effective current (assuming
that the magnetic field oscillation frequency $\omega_B$ is narrow-bandwidth), $\gamma$ is the damping rate for
the target excitations (see section~\ref{secSQLth}),
and $C$ is a constant. 
This expression is valid for integration times
long enough to resolve all of the relevant bandwidths
($t > \Delta m^{-1}, Q_a m_a^{-1}, \omega_1^{-1}, \gamma^{-1}$).
Under fairly general assumptions
defining $\gamma$, we show that $C \le \pi$. 
Under more restricted assumptions, we show
that $C \le 3/2$, and that, if the target is
in a thermal state at temperature $T$,
then $C \le 1/f(n_T)$, where
$f(n_T)$ is a function
of the thermal occupation number at $\omega_1$,
with $f(n_T) \simeq 3/2$ for $n_T \ll 1$ and
$f(n_T) \simeq n_T$ for $n_T \gg 1$.\footnote{
For up-conversion experiments,
the converted power in a narrow frequency band is
at most half of the value from equation~\ref{eqpbound1},
and the ${\rm SNR}^2$ value is $1/4$ of the 
value from equation~\ref{eqsnr2intro}.
}
We conjecture that $C \le 1/f(n_T)$ also holds for our more
general assumptions --- for details, see section~\ref{secSQLth}.
As occurs for the average absorbed power, there is an inverse
relationship between the average SNR and $\Delta m$.

Another common setup has a linear amplifier isolated
from the target, e.g.\ using a circulator
connected to a cold load~\cite{Chapman_2017,Clerk_2010},
to protect the target system from noise. 
In this case, the SNR limit is also given by
equation~\ref{eqsnr2intro}, up to $\OO(1)$ numerical factors.
Perhaps surprisingly, in both of these cases,
 the improved sensitivity
for $\omega_1 \ll m_a$ is actually physical,
and experiments using `down-conversion' in this way
could theoretically achieve improved sensitivities.
However, due to a number of experimental limitations,
including the relatively small $\overline{U}_B$ values obtainable
for high-frequency magnetic fields, realising such
enhancements does not seem to be practical.
Additionally, if $\omega_1 \lesssim L^{-1}$, then the 
EM fields
are naturally in the quasi-static regime,
and the SNR is suppressed by $\sim (\omega_1 L)^2$,
similarly to the absorbed power in equation~\ref{eqpqs1}.

At higher frequencies,
detectors other than amplifiers (e.g.\ photon counters,
bolometric detectors, quasi-particle
detectors, etc.) become easier to implement.
While we could analyse the properties of each
individually, it is the case that for
a wide range of setups, the sensitivity
is bounded by the number of axion quanta absorbed.
We can quantify this using quantum measurement theory.
The Fundamental Quantum Limit (FQL)~\cite{Pang:2019ztr,Tsang:2011zz,Miao:2017vmx} 
for signal detection is determined
by the quantum fluctuations of the EM fields
that couple to the axion signal. Using the arguments 
that lead to equation~\ref{eqpbound1}, we can constrain the
frequency-integrated spectrum of these fluctuations.
For general states, we cannot use 
this to place a bound on detectability,
since there may be cancelling contributions
to this integral.
However, if the sensor interacts
with the target via a damping-type interaction,
e.g.\ an absorptive photodetector or bolometer,
then its effects are equivalent to a passive
load, and the quantum fluctuations of the target EM
fields are the same as in an equilibrium state.
In these circumstances, the sensitivity
to axion DM, over a (fractionally small) mass range
$\Delta m$, is bounded (at the least favourable
axion mass) by
\begin{equation}
	\PP_{\rm det} \le \bar N_a \equiv \frac{\bar P t_{\rm tot}}{\omega_1}
	\label{eqpdet1}
\end{equation}
where $\PP_{\rm det}$ is the probability of detecting
the axion signal.
We will refer to this limit as the PQL (`Passive Quantum
Limit'). It has an obvious interpretation
in terms of photon counting, for schemes
in which axions convert to single photons,
but it also applies to other setups, e.g.\
where a signal consists of multiple quasi-particle excitations.
Coherent-state excitations of the target's EM
fields leave their quantum fluctuations unchanged,
so do not affect the PQL.
As in the SQL case, the $1/\omega_1$ enhancement
for small $\omega_1$ is physical, but probably not practical.

If we take $\gamma$ small enough so that the assumptions
behind equation~\ref{eqsnr2intro} no longer hold,
the maximum SNR from a linear amplifer, isolated behind a circulator
with a cold load,
saturates to $\sim \bar N_a$ (as does the SQL op-amp limit).
This is as we would expect, since the 
target fluctuations are as if we had connected
a passive load.



As we emphasised above, it is certainly possible to do 
better than the PQL, by using techniques involving `non-classical'
EM field states.
One important example 
is using linear amplifiers with correlated
backaction and imprecision noise. By optimising this
correlation, we can in theory obtain
the `quantum limit'~\cite{Clerk_2010,Kampel:2017gze},
for which the SNR bound is
\begin{equation}
	\overline{{\rm SNR}^2} \lesssim 
	\frac{\pi}{2} (g^2 \rho_a \overline{U}_B)^2 \frac{t Q_a}{\omega_1^3 \gamma^2 (1 + n_T)^2}
\end{equation}
where the notation is as for equation~\ref{eqsnr2intro}. Unlike the SQL and PQL,
this limit does not involve $\Delta m$ ---
a QL-limited experiment is inherently broadband, if
we can optimise the amplifier properties across
a wide bandwidth.
In the quasi-static limit, the SNR is again suppressed by 
$\sim (\omega_1 L)^2$. SQUID amplifiers
(as proposed for e.g.\ \cite{Sikivie2014} and the ABRACADABRA axion DM
detection experiment~\cite{Kahn:2016aff}) can,
in some circumstances, attain near-QL performance~\cite{squidhb}.
The fact that the QL-limited sensitivity can, in some
regimes, be better than the PQL, corresponds
to the amplifier back-action
enhancing the quantum fluctuations of the target EM
fields.

While some other measurement schemes, such
as backaction evasion~\cite{Clerk_2010}, do not have such general
limits on their sensitivity, we could still analyse
their performance given more specific assumptions.
In this paper, we will restrict our discussion to the
amplifier and PQL limits introduced above.
One reason for doing so is that, taken together,
they apply to almost all existing
and proposed axion DM detection experiments.

As we will discuss, these limits help
in understanding what can and cannot
enhance an experiment's sensitivity to axion DM,
and in comparing the potential sensitivity of
different kinds of experiments.


\section{Axion DM interactions}
\label{secaxiondmint}

We will suppose that dark matter consists of an axion-like
particle $a$, with a coupling to the SM photon.
This has Lagrangian\footnote{
We take the $(+---)$ signature, and use the convention
$\epsilon_{0123} = -1$. Except where indicated, we use natural units
with $c = \hbar = k_B = 1$.
In general, we will abbreviate $g_{a \gamma \gamma} = g$.}
\begin{align}
	\LL &\supset \frac{1}{2} (\partial_\mu a)^2 - V(a)
-\frac{1}{4} g_{a \gamma \gamma} a F_{\mu \nu} \tilde{F}^{\mu\nu} \nonumber\\
	&=\frac{1}{2} (\partial_\mu a)^2 - V(a) +g_{a \gamma \gamma} a E \cdot B,
\end{align}
where $V(a)$ is the potential for the axion --- in general,
only the mass term $V(a) = \frac{1}{2} m_a^2 a^2$ will be important
for us. 

As far as is understood,
almost all production mechanisms for axion DM result in the field
today being in a coherent, classical-like state~\cite{Kawasaki:2014sqa,Ringwald:2015dsf,Co:2017mop,diCortona:2015ldu,Graham:2015rva};
if the axion mass is small, $m_a \ll \eV$, then the
average occupation number
within the Milky Way is $\gg 1$. 
Since $g_{a \gamma\gamma}$ (and other
couplings) are constrained to be very small, interactions
with a detector will have a negligible effect on the DM's state.
Consequently, for the purposes of
detection, we can treat the DM oscillation as a fixed classical
background field. 

The $F \tilde F$ term is a total derivative,
$F_{\mu \nu} \tilde{F}^{\mu\nu} = 2 \partial_\mu \left(
A_\nu \tilde{F}^{\mu\nu}\right)$, so under integration
by parts, the interaction term in the Lagrangian is equivalent to
\begin{align}
	-\frac{1}{4}g a  F_{\mu \nu} \tilde{F}^{\mu\nu} &\rightarrow \frac{1}{2} g (\partial_\mu a) A_\nu \tilde{F}^{\mu\nu} \\
	&=  \frac{1}{2} g \left(A \cdot (\dot a B + (\nabla a) \times E)  -  A_0 \nabla a \cdot B \right) \\
	&=  - \frac{1}{2} A^\mu J_\mu^{(a)}
	\label{eqLLa}
\end{align}
Note that, for our signature choice, the components of the usual 3-vector potential, which we will denote $A$, are $-A_i$.
To begin with, we will focus on the case
of a spatially-constant (zero-velocity) axion DM field, $\nabla a = 0$.
This is usually a good approximation, since the DM is highly non-relativistic, with
$v_{\rm DM} \sim 10^{-3}$ (we will come
back to the consequences of the axion velocity distribution
in section~\ref{secAxVel}).
If $\nabla a = 0$, then the interaction term
is $\LL \supset \frac{1}{2} g \dot a B \cdot A$.

\subsection{Response dynamics}
\label{secrespdyn}

For many purposes, 
it will be convenient to work in a Hamiltonian
framework. The Lagrangian density for the electromagnetic field can 
be written as 
\begin{equation}
	\LL = - \frac{1}{4} F_{\mu\nu} F^{\mu\nu} - A_\mu J^\mu +
	\LL_{\rm matter}
\end{equation}
where $J_\mu = J^{\rm SM}_\mu + J^{(a)}_\mu/2$.
If we work in Coulomb gauge, $\nabla \cdot A = 0$, 
then for $J_\mu$ independent of $\dot A$ (as
is the case for Dirac fermion matter, and
a zero-velocity axion field), we obtain the Hamiltonian~\cite{Weinberg:1995mt}
\begin{equation}
	H = \int d^3 x \, \left(\frac{1}{2} \dot A^2
	+ \frac{1}{2} B^2 - J \cdot A + \frac{1}{2} J^0 A^0 \right)
	+ H_{\rm matter}
	\label{eqhtot}
\end{equation}
where for Dirac fermion matter, $H_{\rm matter}$ is independent
of $A$. $A^0$ is not a physical degree of freedom,
and can expressed in terms of $J^0$
as $A^0(x,t) = \int d^3x' \frac{J^0(x',t)}{4 \pi |x -x'|} $.


To analyse the effect of the axion oscillation
on the system, we can decompose
the EM vector potential as $A = A_0 + A_1$,
where $A_0 \equiv \langle A \rangle$ 
in the absence of an axion oscillation.
In the notation of appendix~\ref{appQmeas}, 
$A_1 = \Delta A$. If we assume that $A_1$ is small
compared to $A_0$, then 
\begin{align}
	A \cdot B&= 
	A \cdot (\nabla \times A)  \\
	&\simeq A_0 \cdot (\nabla \times A_0) + A_1 \cdot (\nabla \times A_0) \nonumber \\
	&\,\,\,\,\,\, + A_0 \cdot (\nabla \times A_1) 
\end{align}
so the axion interaction term can be expanded as 
\begin{equation}
	A \cdot B \simeq A_0 \cdot B_0 + 2 A_1 \cdot B_0 + \nabla \cdot (A_1 \times A_0)
\end{equation}

In general, $B_0$ will have some time dependence.
To start with, we will assume that the time depedence
and spatial profile factorise, so $B_0 = B_0(t) b(x)$,
as is the case
for e.g.\ a cavity standing mode (we will revisit
this in section~\ref{secAxVel}).
We can decompose $A_1 = A_b b + A_\perp$,
where $\int dV A_\perp \cdot b = 0$;
we will also write $A_0 = A_b^{(0)} b + A^{(0)}_\perp$, etc.
Then, writing $V_b \equiv \int dV b^2$, the $A_b$-dependent
parts of the Hamiltonian are
\begin{align}
	H &= V_b \left(\dot A_b^{(0)} \dot A_b + \frac{1}{2} \dot A_b^2 
	- g \dot a A_b B_0 \right)\\
	&+ \int dV \left(\frac{1}{2} B^2 - J_{\rm SM} \cdot A\right) + \dots
	\label{eqhab1}
\end{align}
The first two terms are the only ones depending on $\dot A_b$.
The conjugate momentum to $A$ is $E$, so
the conjugate momentum to $A_b$ is $- V_b E_b = V_b \dot A_b$
(since $\int dV (\nabla A_0) \cdot b = - \int dV A_0 \nabla \cdot b = 0$), with equal-time commutation relation $[\hat A_b, \hat E_b] = - i/V_b$.
Consequently, the Hamiltonian for $A_b$ is analogous to
that for a 1D oscillator,
\begin{equation}
	\hat H_{\rm 1D} = \frac{(\hat p + p_0)^2}{2 M}
	- g j(t) \hat x + V_{\rm int}(\hat x, \dots)
\end{equation}
where $\hat x \equiv A_b$, $\hat p \equiv -E_b V_b$,
$p_0 \equiv -E_b^{(0)} V_b $,
$j(t) \equiv \dot a (t) B_0(t) V_b$,
and $M \equiv V_b$. Here, $V_{\rm int}(\hat x)$ summarises
the other terms in equation~\ref{eqhab1}, which do not 
depend on $\hat p$.

If we consider a very short $j(t)$ pulse,
turning on and off much faster than the system's dynamics,
then its effect is to impulsively change
$p$ by $g \int dt \, j(t) \equiv g J$.
Averaging over possible signs of the pulse,
the expected energy absorbed is
$\langle W \rangle \simeq (\Delta p)^2 / (2 M)$.
In our case, since $j$ depends on the time derivative
of $a$, a delta-function
$j$ pulse corresponds to a step function in
$a(t)$,
and we have $\langle W \rangle \simeq \frac{1}{2}
g^2 (\Delta a)^2 V_b B_0^2$.

This argument tells us the expected energy absorbed
by the target from a very fast axion
field `pulse'.
However, as discussed above, we expect
axion DM to be a narrow-bandwidth oscillation,
with fractional bandwidth $\sim 10^{-6}$.
If $g$ is small enough that the target is in the linear
response regime, then the energy it absorbs
from a finite-time $j_t(t)$ signal is 
\begin{equation}
	\langle W \rangle = \frac{g^2}{2 \pi} \intinf d\omega \, \omega  \,
	|\tilde j_t(\omega)|^2 
	{\rm Im} \tilde \chi (\omega)
\end{equation}
where $\chi$ is the linear response function for $x$
(if the dynamics are non-stationary in time, we can
consider averaging over all possible starting times).
A delta-function pulse has equal power at all frequencies,
so
\begin{equation}
	\langle W \rangle = \frac{g^2}{2 \pi} J^2 \intinf
	d\omega \, \omega  \,
	{\rm Im} \tilde \chi (\omega)
\end{equation}
Equating this to the energy $g^2 J^2 / (2 M)$ absorbed
from the pulse, we have
\begin{equation}
	\intinf d\omega \, \omega \, {\rm Im}\tilde\chi(\omega)
	= \frac{\pi}{V_b}
	\label{eqchisum}
\end{equation}
This `sum rule' is analogous to the Thomas-Reiche-Kuhn sum rule
for `oscillator strengths' in atomic physics~\cite{Sakurai}.

By itself, equation~\ref{eqchisum} does give
us any limit on the response in a specific
frequency range,
since the integrand could have large cancelling components.
However, if $\omega \, {\rm Im} \tilde \chi (\omega)$
is always $\ge 0$, then we can bound the absorbed power
from any signal. 
This obviously applies
if the target is in its ground state
(since a forcing can only add energy), or
if ${\rm Im} \tilde \chi$ is equivalent to its
ground-state form. 
For a purely harmonic oscillator, the latter is true
in any state.
More generally, if the target
is in a mixed state, where the probability of a microstate
decreases with increasing energy, then the condition
also holds, as we show below.

These arguments are a generalisation of the pulse-absorption argument
from~\cite{Baryakhtar:2018doz}, which was used
to the determine the axion-mass-averaged signal power
from a dielectric haloscope. 

The sum rule in equation~\ref{eqchisum} depends on the fact
that laboratory materials are composed of particles
carrying electric charges. 
If particles carrying magnetic charge existed
(or more generally, particles with Amperian
electric dipole moments), then it would be possible
to violate the sum rule. This can be seen using
a electric-magnetic duality transformation, under which a paramagnetic material
is transformed to a material with negative
electrical susceptibility. As we go over below,
a dielectric medium gives $\int d\omega \, \omega \, {\rm Im} \tilde
\chi(\omega) = \frac{\pi}{\epsilon V_b}$, so if
$\epsilon < 1$ over some order-1 frequency range, 
the sum rule can be violated.
This illustrates that the sum rule does have physical
content, but is satisfied very generally for
realistic laboratory systems.

\subsection{Fluctuation sum rules}

To apply the FQL detectability limits discussed
in appendix~\ref{appQmeas}, we
need to understand the fluctuation spectrum
of $A_b$. We can relate this to the response
function via the Kubo formula~\cite{tongKinetic},
${\rm Im}\tilde \chi(\omega)
= \frac{1}{2} (S_{A_b A_b}(\omega) - S_{A_b A_b}(-\omega))$,
where $S_{A_b A_b}(\omega)$ is the spectral
density of $A_b$ fluctuations.
Thus, the sum rule in equation~\ref{eqchisum}
implies a corresponding sum rule for $S_{A_b A_b}$,
\begin{equation}
	\intinf d\omega \, \omega \, S_{A_b A_b}(\omega)
	= \frac{\pi}{V_b}
	\label{eqSum1}
\end{equation}
We can also derive this sum rule directly from the
commutation relations of the EM fields.
The spectral density of $A_b$ fluctuations (assuming that
they are stationary in time) is 
\begin{equation}
	S_{A_b A_b}(\omega) \equiv \intinf dt\, e^{i \omega t} \langle \hat A_b(t) \hat A_b(0) = \mathcal{F} \langle \hat A_b(t) \hat A_b(0) \rangle
\end{equation}
where $\hat A_b(t)$ is the Heisenberg picture operator
for the system, in the absence of axion interactions (going forwards,
we will drop the hats). 
Integrating this over $\omega$,
\begin{align}
	\int_{-\infty}^{\infty} d\omega \, \omega \, S_{A_b A_b}(\omega) &= 
	\int_{-\infty}^{\infty} d\omega \, \omega \, \FF \langle A_b(t) A_b(0) \rangle \\
	 &= i \int_{-\infty}^{\infty} d\omega \, \FF \langle \dot{A_b}(t) A_b(0) \rangle
\end{align}
We have
\begin{equation}
	E_b 
	= \frac{1}{V_b} \int dV \, E \cdot b  
	= \frac{1}{V_b} \int dV \, (- \nabla A_0 - \dot A) \cdot b
	= - \dot A_b
\end{equation}
since $\nabla \cdot b = 0$, so
\begin{equation}
	\int_{-\infty}^{\infty} d\omega \, \omega \, S_{A_b A_b}(\omega) 
	= - 2 \pi i \langle E_b(0) A_b(0) \rangle
	\label{eqeb0ab0}
\end{equation}
If the fluctuations of
$A_b$ are stationary in time (as for e.g.\ a coherent state), 
then 
$\FF \langle A_b(t) A_b(0) \rangle$ is real.
So, $\langle A_b(0) E_b(0) \rangle$ is imaginary,
and consequently, for equal times,
\begin{equation}
	\langle E_b A_b \rangle = \frac{1}{2}\langle [E_b,A_b] \rangle
	= \frac{i}{2 V_b}
\end{equation}
Inserting this into equation~\ref{eqeb0ab0}
reproduces equation~\ref{eqSum1}.

As per the previous section, we are usually interested
in the fluctuations across some narrow frequency range.
In general, there can be contributions to equation~\ref{eqSum1}
from positive and negative $\omega$,
leading to cancellations. However, for the ground
state of the system, $S_{A_b A_b}(\omega) = 0$
for $\omega < 0$; for any operator $\hat F$, 
\begin{equation}
	\langle 0 | \hat F(t) \hat F(0) | 0 \rangle
	= \sum_n e^{- i (\omega_n - \omega_0) t} |\langle n | \hat F | 0 \rangle|^2
\end{equation}
So, for the ground state, we obtain the sum rule
\begin{equation}
	\int_0^\infty d\omega \, \omega \, S_{A_b A_b}(\omega) = \frac{\pi}{V_b}
	\label{eqssum1}
\end{equation}
The same is true for coherent states, since their fluctuations on top
of the c-number expectation value are the same as for the ground state.

If we consider a probability mixture of energy
eigenstates, $\rho = \sum_n p_n |n\rangle \langle n|$, then
\begin{align}
	&\langle \hat F(t) \hat F(0) \rangle
	= \sum_m p_m \sum_n e^{-i (\omega_n - \omega_m) t}
	|\langle n | \hat F | m \rangle|^2 \nonumber \\
	&= \sum_m p_m |\langle m | \hat F | m \rangle|^2 +  \sum_m \sum_{n > m} 
	|\langle n | \hat F | m \rangle|^2 \times \nonumber \\
	&\quad\left(
	p_m e^{-i (\omega_n - \omega_m) t} + p_n e^{-i (\omega_m - \omega_n) t} \right) \nonumber \\
	&= \sum_m p_m |\langle m | \hat F | m \rangle|^2 +  \sum_m \sum_{n > m} 
	|\langle n | \hat F | m \rangle|^2 \times \nonumber \\ 
	& \quad\left(
	(p_m - p_n) e^{-i (\omega_n - \omega_m) t} + 2 p_n \cos ((\omega_m - \omega_n) t)\right)
\end{align}
Consequently, if $p_m \ge p_n$ for $m < n$, then since
the odd part of this expression determines
${\rm Im} \tilde \chi(\omega)$, we have
${\rm Im} \tilde \chi (\omega) \ge 0$ for $\omega \ge 0$.
Thus, for mixed states such as thermal states, the
sum rule from equation~\ref{eqchisum} applies.

While we have focussed on the fluctuations
of $A_b$, the full interaction operator
with the axion field is $- \frac{1}{2} g \dot a A \cdot B$.
(equation~\ref{eqhtot}). In most of this paper, we assume that 
there is a large `background' field $B_0$,
which is basically classical, in the sense
that the fluctuations around its expectation value
are fractionally small.
Then, as derived in Section~\ref{secrespdyn},
the interaction operator is approximately 
$- g \dot a B_0 A_b$, so the fluctuations of
$A_b$ determine the FQL for detecting an axion
forcing. 

\subsection{Effective Hamiltonians}

The above derivations relied on the conjugate momentum
of $A_i$ being $E_i$, i.e.\ there being no
other terms in the Hamiltonian involving $\dot A_i$. 
For example, if we were considering a dielectric
medium, where the energy density is $\epsilon E^2$, then
the conjugate momentum to $A_i$ would be $\epsilon E_i$, 
and we would have
$\int d\omega \, \omega S(\omega) = \frac{\pi}{\epsilon V_b}$.
Thus, if e.g.\ a resonant cavity is filled with dielectric
material, the power it is able to absorb decreases~\cite{sclong}.
From above, we know that once all of the dynamics are
taken in account,
$\intinf d\omega \, \omega S(\omega) = \frac{\pi}{V_b}$.
This implies that the `extra' fluctuations must be at frequencies
above the validity of the effective Hamiltonian.

Similarly, the $\pm \infty$ limits of the $\omega$ integrals
above should not be taken literally --- at the very least, electroweak physics
arises at some energy scale! What we can infer is that,
for frequency ranges over which our description
of the system is good,
$\int d\omega \, \omega \, {\rm Im} \tilde \chi
\le \frac{\pi}{V_b}$ in the ground state, and so on.

\subsection{Axion velocity}
\label{secAxVel}

So far, we have taken the axion velocity to be zero.
This will not be strictly true; in most
particle physics models, axion DM in the galaxy is expected
to have a virialized velocity distribution, with typical
velocity $\sim 10^{-3}$~\cite{Necib:2018iwb}.
There may also be components with smaller velocity
dispersions, arising from either non-virialized
`streams' of dark matter particles~\cite{Sikivie1992,Freese2004,Duffy2008,Vogelsberger2011}, or from
bound `mini-halos' 
(see e.g.~\cite{Fairbairn2018,1909.11665}).
In most of our discussions, we will assume that
the DM velocity distribution is characterised by
a single velocity dispersion scale, which
is taken to be the virial velocity spread.
Extending to more complicated velocity distributions
is straightforward.\footnote{Sufficiently small mini-halos can also transit the detector on short timescales,
resulting in a strongly time-dependent
axion DM density \cite{1909.11665}; we defer discussion
of this case to future work.}

As per equation~\ref{eqLLa}, the interaction term
is $\LL \supset - \frac{1}{2} A^\mu J_\mu^{(a)}$, where 
\begin{equation}
	J^{\mu}_{(a)} = g \begin{pmatrix}
		(\nabla a) \cdot B \\
		\dot a B + (\nabla a ) \times E
	\end{pmatrix}
\end{equation}
Compared to the zero-velocity case, the axion velocity term
$\nabla a$ results in a coupling to the scalar potential
$A_0$, as well as the vector potential $A$.
However, we can work in a gauge in which $A_0 = 0$, in which case
the extra coupling term is 
\begin{equation}
	\LL \supset A_1 \cdot ((\nabla a) \times E_0)
\end{equation}
(after integration by parts). For an axion wave of definite
momentum, this corresponds to the $B$ field in the axion rest frame,
as expected.

Consequently, we can replace $\dot a B_0$ by $\dot a B_0 + (\nabla a)
\times E_0$ as our forcing term. Since
$|\nabla a| \sim v_a |\dot a| \sim 10^{-3} |\dot a|$,
and
the attainable (static) magnetic
fields in laboratories are significantly larger than
attainable electric fields, the $\dot a B_0$ term dominates in almost all circumstances of interest.

A more important effect of the axion velocity distribution
is that the axion signal is no longer a spatially uniform, single-frequency
oscillation. If the experimental volume is significantly
smaller than the axion coherence length ($L_{\rm coh} \sim (v_a m_a)^{-1} \sim 10^3 m_a^{-1}$),
then the axion field inside the volume is approximately
uniform, but is incoherent over times
$\gtrsim v_a^{-2} m_a^{-1}$, corresponding to a frequency spread $\sim v_a^2 \nu_a \sim 10^{-6} \nu_a$ (where $\nu_a \equiv m_a/(2 \pi)$
is the axion frequency). If the experimental volume is larger, then 
the axion field is incoherent over distances
$\sim v_a^{-1} m_a^{-1}$.

We can treat the spatial variation of the axion field,
as well as any time-dependence of the $B_0$ spatial profile,
by decomposing $\dot a(t) B_0(t)$ into spatially orthogonal modes,
each with their own time dependence. Writing
\begin{equation}
	\dot a(t) B_0(t) = \sum_i \dot a_{b_i}(t)  B_{b_i}(t) b_i
\end{equation}
where $\int dV b_i \cdot b_j = \delta_{ij} V_{b_i}$,
and $a_{b_i}$ and $B_{b_i}$ are functions only of time,
we have
\begin{equation}
	H_{\rm int} = g \sum_i \dot a_{b_i} B_{b_i} A_{b_i}
\end{equation}
We have equal-time commutation relations
\begin{equation}
	[A_{b_i}, E_{b_j}] = - \delta_{ij} \frac{i}{V_{b_i}}
\end{equation}
So, the cross terms in the sum rule (eq.~\ref{eqssum1}) vanish, giving
\begin{equation}
	\int_{-\infty}^{\infty} d\omega \, \omega \, \FF \langle A_{b_i}(t_0) A_{b_j}(t_0 + t) \rangle
	= \delta_{ij} \frac{\pi}{V_{b_i}}
	\label{eqdijsum}
\end{equation}
This is as we would expect from the impulse
argument above --- over very short timescales,
there is no dynamics coupling
the spatially orthogonal target modes,
so they have independent responses to pulses.
We will discuss some of the consequences of this
in section~\ref{secSQLth}.


\section{Parametrics of DM detection}
\label{secTheorySens}

The sum rules (equations~\ref{eqchisum}, \ref{eqssum1}, \ref{eqdijsum}) derived in Section~\ref{secaxiondmint}
can be used to bound the average power absorbed
from the axion effective current, in an axion
DM detection experiment.\footnote{In
most of this paper, we view the 
magnetic field $B_0$ as a classical background field,
and consider at the energy absorbed from the axion
effective current $J^{(a)} = g \dot a B_0$. For a 
static $B_0$, this is equivalent to the energy absorbed
from the axion field by the SM system. For time-dependent
$B_0$, the energy absorbed `from $J^{(a)}$'
by the rest of the SM system includes energy transferred
from the magnetic field. While we could always work with
the energy absorbed from the axion field itself,
the detectability of small signals is more closely related to
the energy absorbed from $J^{(a)}$.}
The simplest case is when the $B_0$ field is static.
Over sufficiently long integration times,
so that we resolve the spectral
features of the axion signal,
the expected time-averaged power absorbed
for an axion of mass $m$ is
\begin{equation}
	P_m \simeq \frac{g^2 B_0^2 V_b^2}{2 \pi} \intinf d\omega \, 
	\omega S_{\dot a \dot a}(\omega) \tchi_i (\omega)
	\label{eqpm1}
\end{equation}
where we write $\tchi_i \equiv {\rm Im} \, \tchi$.\footnote{for shorter integration times $t_{\rm int}$, $S_{\dot a \dot a}$
should be convolved with a kernel of width $\sim 1/t_{\rm int}$.}
Averaging this over different axion masses $m$, we
can use the fact that, since the axion bandwidth
is small, $\delta \omega_a \sim 10^{-6} m$, integrating
over $m$ for fixed $\omega$ is approximately
the same as integrating over $\omega$ for fixed $m$, 
\begin{equation}
	\intoinf dm \, S_{\dot a \dot a}(\omega) 
	\simeq \intoinf d\omega' \, S_{\dot a \dot a} (\omega')|_{m = \omega} \simeq \pi \rho_a
	\label{eqsintinf}
\end{equation}
Hence, 
\begin{align}
	\intoinf dm \, P_m &\simeq \frac{g^2 B_0^2 V_b^2 \rho_a}{2} \intinf d\omega \, \omega \, \tchi_i (\omega) \\
	&= g^2 \rho_a V_b B_0^2 \frac{\pi}{2}
	\label{eqpstatic}
\end{align}
Consequently, the absorbed power, integrated over all
axion masses, is set by the magnetic field
energy in the $B_0$ field (ignoring magnetic
fields on very small spatial scales
--- see appendix~\ref{apmag}).

If we are interested in looking for an axion
within a specific mass range $\Delta m$, then
we can average $P_m$ over that mass range,
\begin{equation}
	\bar P \equiv \frac{1}{\Delta m} \int_{\Delta m}
	dm \, P_m \lesssim
\frac{g^2 \rho_a V_b B_0^2}{\Delta m} \frac{\pi}{2}
	\label{eqpbar1}
\end{equation}
where the inequality assumes that $\omega \, \tchi_i (\omega) \ge 0$ for all $\omega$.
Equality can be obtained if the response function is concentrated
into the
$\Delta m$ range (we will discuss 
some of the experimental practicalities
of this in section~\ref{secdmdetect}).
Since $\min_{m \in \Delta m} P_m \le \bar P$, the
smallest absorbed power for any axion mass
within the range $\Delta m$ is upper-bounded
by equation~\ref{eqpbar1}. 
As expected, searching over a smaller axion
mass range permits higher conversion
powers within that range.

Equation~\ref{eqpbar1} applies if $\tchi_i(\omega) \ge 0$
for all $\omega > 0$. Even if this is not the case,
it can also apply if $\tchi_i(\omega) = \tchi_i'(\omega)$
for $\omega$ in the mass range $\Delta m$, where $\tchi_i'$
is the response function for a system which
\emph{does} have $\tchi_i(\omega) \ge 0$ for all $\omega > 0$.
Similarly, if $\tchi_i \le C \tchi_i'$ in the relevant
frequency range, for some constant $C$, then the bound
in equation~\ref{eqpbar1} should be multiplied by $C$.
For example, if a system has has $\tchi_i(\omega) < 0$
at high frequencies, but has simple low-frequency behaviour,
these considerations can be useful.

In many cases, instead of operating a single
experimental configuration for
the whole observation time,
we `tune' our experiment by operating it in different
configurations, one after the other.
The average power for a given axion mass is
the appropriately weighted sum of the powers
from the different configurations, and the
corresponding limits apply.

The equations above apply to the whole experimental
apparatus. However, a common experimental setup
is to have a conductive shield (e.g.\ an EM cavity)
inside a larger magnetic field. If the relevant
dynamics inside and outside the cavity are independent,
then we can apply the above arguments to the volume
inside the cavity, replacing the total magnetic field
energy by the energy inside the cavity.

\subsection{PQL limits}
\label{secdetect}

From appendix~\ref{appQmeas}, if the quantum fluctuations of $A_b$
are stationary, then the $\OO(g^2)$ formula for the probability of the axion
interaction changing the state of the target system
is 
\begin{equation}
	\PP_{\rm ex} \simeq \frac{g^2 B_0^2 V_b^2 t_{\rm exp}}{\pi}
	\intoinf d\omega \, S_{\dot a \dot a} (\omega) \SC_{A_b A_b}(\omega)
\end{equation}
where $\bar{S}$ denotes the symmetrised spectral density, and
we assume that $t_{\rm exp}$ is much longer
than the inverse bandwidth of spectral features.
If the fluctuations are equal to those in the ground
state, then $\SC_{A_b A_b}(\omega) = \tchi_i (\omega)$ for $\omega > 0$. In that case,
\begin{equation}
	\PP_{\rm ex} \simeq \frac{g^2 B_0^2 V_b^2 t_{\rm exp}}{\pi}
	\intoinf d\omega \, S_{\dot a \dot a} (\omega) 
	\tchi_i (\omega)
	\simeq \frac{P_m}{m}
\end{equation}
where the latter equality holds since $S_{\dot a \dot a}$ is tightly concentrated around
$m$. Hence, $\PP_{\rm ex} \simeq N_a$, the expected number
of quanta absorbed. Since $\tchi_i$ for the ground
state satisfies $\tchi_i(\omega) \ge 0$ for $\omega > 0$, 
equation~\ref{eqpbar1} holds, and so
the axion-mass-averaged excitation
probability satisfies
\begin{equation}
	\bar \PP_{\rm ex} \lesssim
	\frac{g^2 \rho_a V_b B_0^2 t_{\rm exp}}{m \Delta m} \frac{\pi}{2}
	\equiv \bar N_a
	\label{eqppex1}
\end{equation}
where the average is taken over a fractionally-small
axion mass range $\Delta m$, centred on $m$.
To be confident of identifying or excluding an axion
signal, we need $\bar N_a \gtrsim$ few.
More generally, if $\SC_{A_b A_b}$ for the operational
state of the detector satisfies 
\begin{equation}
\int_{\Delta m} d\omega \, \omega \SC_{A_b A_b} \lesssim
	\frac{\pi}{2 V_b}
	\label{eqscb1}
\end{equation}
then equation~\ref{eqppex1} also holds.

In section~\ref{secPQL}, we discuss the circumstances under
which the limit in equation~\ref{eqppex1} applies, and can be 
achieved. The most obvious example,
in which this limit can be achievable, is the case
of an absorptive, background-free photon counter.
In the presence of noise sources
(such as thermal noise or detector noise),
it may not be possible to attain this limit.
Conversely, if a detection setup does
not satisfy equation~\ref{eqscb1}, 
then the FQL still places limits on its sensitivity,
but these will depend on how much the fluctuations
exceed the PQL value.

\subsection{SQL op-amp}
\label{secSQLth}

If we read out our signal using 
a phase-invariant, SQL-limited amplifier
coupled weakly to the target (i.e.\ in `op-amp'
mode~\cite{Clerk_2010}), then from
appendix~\ref{seclinamp}, the
SNR from an axion signal satisfies
\begin{equation}
	{\rm SNR}^2 \le (g B_0 V_b)^4 t \intoinf 
	\frac{d\omega}{2\pi} \left(\frac{S_{\dot a \dot a} |\tilde \chi|^2}{|\tilde \chi| + \tchi_i + S_n}\right)^2
	\label{eqsqll1}
\end{equation}
where $S_n$ summarises the effects of any additional
noise (beyond amplifier backaction, imprecision,
and zero-point fluctuations),
referred back to $A_b$. 
For example, if the target is subject to thermal noise at temperature $T$, we have $S_n \ge
2 n_T \tchi_i$,
where
$n_T(\omega) \equiv (e^{\omega/T}-1)^{-1}$.\footnote{
	While the thermal fluctuations of
	$A_b$ are set by $2 n_T \tchi_i$,
	if the amplifier
	is coupled to other degrees of freedom, then
	the total effect of thermal noise
	on the output may be greater.
	}
We assume that the spatial profile
of $\Ja$ can be treated as constant in time, to begin with.

If we are interested in a mass range
$\Delta m \gg \delta \omega_a$ (i.e.\ a fractional
mass range $\gg 10^{-6}$), then the
quantity determining the axion-mass-averaged
${\rm SNR}$ squared is
\begin{equation}
	S \equiv \int_{\Delta m} d\omega |\tchi|^2 
	\left(\frac{1}{1 + (1 + 2 n_T) \tilde \chi_i /|\tilde \chi|}\right)^2
	\label{eqs1}
\end{equation}
The simplest form of response function is a
single-pole oscillator, which has
$\tchi(\omega) = \frac{-1/M}{\omega^2 - \omega_0^2
+ i \omega \gamma}$, where $\omega_0$ is the
undamped frequency, and $\gamma$ is the damping
rate. Evaluating $S$ for this $\tchi$, we find
that if $\Delta m \ge (1 + 2 n_T) \gamma$, the integral
is dominated by a bandwidth $\sim (1 + 2 n_T) \gamma$, 
and we have
\begin{equation}
	S \le \frac{1}{2 V_b^2 m^2 \gamma f(n_T)}
	\label{eqssp}
\end{equation}
where 
\begin{equation}
	f(n_T) \simeq \begin{cases}
		\frac{3}{4} & n_T \ll 1 \\
		n_T & n_T \gg 1
	\end{cases}
\end{equation}
assuming that $\gamma \ll m$.

The dependence of eq.~\ref{eqssp} on $\gamma$ illustrates that
the limit on $S$ will depend on what
we assume about the damping properties of our system.
As a simple example, suppose that an $A_b$ oscillation
of amplitude $C$, for which the electric field
energy is $U_E =\frac{1}{2} M \omega^2 |C|^2$, results
in a dissipated power of $P_{\rm diss} \ge \gamma U_E$,
for some $\gamma$
(for example, due to resistivity in the walls of a resonant
cavity). The $A_b$ response to a monochromatic axion
oscillation has $|C|^2 = |\tchi|^2 |j|^2$, and
the cycle-averaged absorbed power from the axion
effective current is $P_{\rm abs} = \frac{1}{2}|j|^2 \omega \tchi_i$.
If the target is a passive system, then $P_{\rm abs}
\ge P_{\rm diss}$, so we must have $\tchi_i \ge M \omega \gamma 
|\tchi|^2$. As we would expect, a single-pole
response function with damping rate $\gamma$ has
$\tchi_i = M \omega \gamma |\tchi|^2$.

We can immediately use
the $\tchi_i \ge M \omega \gamma |\tchi|^2$ condition
to place a limit on $S$,
since 
\begin{equation}
	S \le \int_{\Delta m} d\omega |\tchi|^2 
	\le \int_{\Delta m} d\omega \frac{\chi_i}{M \omega \gamma}
	\lesssim \frac{\pi}{2 V_b^2 m^2 \gamma}
	\label{eqsboundgen}
\end{equation}
This limit is only $\OO(1)$ larger than the $n_T = 0$
single-pole result (equation~\ref{eqssp}).
In fact, we conjecture that a single-pole response
function maximises $S$, subject to the 
$\tchi_i \ge M \omega \gamma |\tchi|^2$ condition.
With more restrictive conditions on $\tchi$, 
we can prove this. For example, if $\tchi$ can be written
as a sum (with positive coefficients) of single-pole response functions with damping
rate $\ge \gamma$, then since
the integrand of $S$ is convex in the $(\tchi_r, \tchi_i)$
plane, $S$ is less than or equal to the sum
of the integrals for the single-pole response functions.
This form of response function is applicable when e.g.\ 
the dissipation rate for all of the relevant modes
of the target is $\gamma$ (as would arise if we
coupled together oscillators with damping rate $\gamma$).

To relate $S$ to the SNR for an axion DM signal, 
we can for simplicity take the axion signal 
to have a top-hat spectral
form, 
$S_{\dot a \dot a} \simeq 
\frac{\pi \rho_a}{\delta \omega_a} \mathbf{1}_{\delta \omega_a} = \frac{\pi \rho_a Q_a}{m} \mathbf{1}_{\delta \omega_a}$.
In this case,
\begin{align}
	{\rm SNR}^2 &\lesssim (g^2 B_0^2 V_b^2 \rho_a)^2  
	\frac{t \pi Q_a^2}{2 m^2} \times \\ &\int_{\delta \omega_a} 
	d\omega |\tchi|^2 
\left(\frac{1}{1 + (1 + 2 n_T) \tilde \chi_i /|\tilde \chi|}\right)^2
\end{align}
So, averaging this over the $\Delta m$ axion mass range,
\begin{equation}
	\overline{{\rm SNR}^2} \lesssim \frac{\pi}{2} (g^2 B_0^2 V_b \rho_a)^2
	\frac{t Q_a}{m^3 \Delta m \gamma} 
	\left(\frac{S}{(V_b^2 m^2 \gamma)^{-1}}\right)
\end{equation}
Using equation~\ref{eqsboundgen},
this gives
\begin{equation}
	\overline{{\rm SNR}^2} \lesssim \frac{\pi^2}{4} (g^2 B_0^2 V_b \rho_a)^2
	\frac{t Q_a}{m^3 \Delta m \gamma} 
	\label{eqsnr22}
\end{equation}
If $S$ is maximised by a single-pole response, then 
\begin{equation}
	\overline{{\rm SNR}^2} \lesssim \frac{\pi}{4} (g^2 B_0^2 V_b \rho_a)^2
	\frac{t Q_a}{m^3 \Delta m \gamma f(n_T)} 
	\label{eqsnr22sp}
\end{equation}
Writing this in terms of the $\bar P$ expression
from above, 
\begin{equation}
	\overline{{\rm SNR}^2} \lesssim 
	\frac{\bar P_{\rm max}^2 Q_a t \Delta m}{\pi m^3 \gamma f(n_T)}
	= \frac{\bar P_{\rm max}^2 Q_a Q t \Delta m}{\pi m^4 f(n_T)}
	\label{eqsnr2pbar}
\end{equation}
where $Q \equiv \gamma / m$, and
$\bar P_{\rm max}$ is the quantity from equation~\ref{eqpbar1}.
As mentioned above, for each resonant configuration,
the bandwidth contributing most of the ${\rm SNR}^2$ is
$\sim (1 + 2 n_T) \gamma$, so we need to scan over
multiple different configurations to have
sensitivity over a wide mass range. Conversely,
if $n_T$ is large enough that $\Delta m \lesssim 2 n_T \gamma$,
then a single resonant configuration
can cover the entire mass range approximately equally,
giving
\begin{equation}
	{\rm SNR}^2 \lesssim
	\frac{\pi}{8} (g^2 B_0^2 V_b \rho_a)^2 \frac{t Q_a}{m^3 \gamma^2 n_T^2}
	\label{eqsnr2bband}
\end{equation}

It should be noted that, to achieve the SNR bound
in equation~\ref{eqsqll1}, which assumes
minimum added noise $\bar S_{A_b A_B}^{\rm add} = |\tilde \chi|$
(see appendix~\ref{appQmeas}), it is necessary for
the backaction spectral density to be $\bar S_{FF} = 1/(2|\tchi|)$
at all $\omega$. If we had an amplifier connected
directly to e.g.\ a high quality factor
cavity, this would require its noise impedance 
to vary rapidly as a function of frequency. However,
if we take $\bar S_{FF} \simeq {\rm const}$, then
by choosing this constant value appropriately,
we can do $\OO(1)$ as well. For example, with a narrow-bandwidth single pole
response function, the optimum constant value of 
of $\bar S_{FF}$ for $n_T = 0$ is $\simeq M \omega \gamma$
(which is twice the optimum value
on-resonance), giving $S$ a factor $2\pi/9 \simeq 0.7$ smaller
than choosing the optimum $\bar S_{FF}$ at every frequency.

Another point to note is that 
it may be possible to place tighter bounds
on $S$ by analysing the system's dissipation more
carefully. In particular, if the system's response
to an axion forcing
has stored energy $U > U_E$, and $P_{\rm diss}
\ge \gamma U$, then $P_{\rm abs} \ge P_{\rm diss}$
implies that
\begin{equation}
	\tchi_i \ge \frac{U}{U_E} M \omega \gamma |\tchi|^2
	\label{equue}
\end{equation}
and in effect, one can replace $\gamma$ by $\gamma U / U_E$
in the expressions above.
We discuss an example of this in section~\ref{secQuasiStatic},
where we consider oscillations in the quasi-static
regime, for which the stored magnetic field energy
is much larger than the energy in the in electric fields.

The above expressions are only valid when
the relevant integration times are long compared
to the inverse bandwidth of spectral features.
Looking at the $n_T \ll 1$ limit,
if we consider a single-pole resonator with $Q \gg Q_a$,
then we need at least $\Delta m / \delta \omega_a$
tuned configurations to have sensitivity across
the $\Delta m$ mass range, so we can
spend at most $t_1 = t_{\rm tot} \frac{\delta \omega_a}{\Delta m}$ in each of them. If $t_1 \lesssim Q/\nu$, then
we do not resolve the resonator bandwidth,
and the formulae above do not hold.
In particular, if we fix $t_{\rm tot}$, then
for $Q \gtrsim \nu t_1$, we do not expect the SNR
to continue improving with increasing $Q$
(this limiting $Q$ is often large, but may sometimes
be of practical relevance~\cite{srf}).
Plugging $Q \simeq \nu t_1$ into equation~\ref{eqsnr2pbar},
which should be parametrically (though not numerically)
valid, we obtain
\begin{equation}
	\overline{{\rm SNR}^2} \lesssim 
	\frac{1}{3 \pi^2} \left(\frac{\bar P_{\rm max} t_{\rm tot}}{m}\right)^2 \simeq \left(0.2 \bar N_{\rm max}\right)^2
	\label{eqSQLn1}
\end{equation}
so the SNR limit is parametrically the same
as the PQL. 
In simple cases, such as when the back-action
power spectral density is frequency-indepenent,
this corresponds to the quantum fluctuations of $A_b$
satisfying the PQL sum rule (at least to $\OO(1)$).\footnote{In the framework of~\cite{Clerk_2010}, the `fluctuations' of $A_b$ can be
made large by increasing the amplifier-target coupling.
However, this also increases the amount we learn
about $A_b$ from the amplifier output; to determine
the \emph{quantum} fluctuations of $A_b$, we need to condition
on the observed output of the amplifier
(c.f.\ the analysis of backaction evasion in~\cite{Clerk_2008}).}

If the spatial profile of $\Ja$ is not constant,
then as discussed in section~\ref{secAxVel}, 
we can consider the time variation
in an orthogonal set of spatial profiles.
This is necessary if the axion
coherence length is smaller
than the scale of the experiment
(or the extent of the background magnetic field,
whichever is smaller),
i.e\ if $\nu_a \gtrsim (v_a L)^{-1}
\sim 300 \GHz \, ({\rm meter} /L)$.\footnote{In contrast to how
the fluctuation spectrum can be concentrated into a narrow-bandwidth
peak, at the expense of surrounding frequencies, the
frequency-averaged fluctuations for each orthogonal spatial
mode are fixed by its spatial profile, and cannot
be concentrated into one spatial mode at the expense of others.
In the impulse picture, 
if we imagine a set of simultaneous impulses at different spatial points, then
causality prevents the energy absorbed from depending
on their relative signs etc, whereas the reponses
to impulses at different times (picking out a specific frequency)
can interfere with each other.
}
The SNRs from these orthogonal spatial
profiles will add in quadrature.

\subsection{Isolated amplifier}
\label{secIsolated}

For the op-amp coupling considered in the previous
section, the amplifier's backaction noise
couples to $A_b$.
Another way to couple an amplifier to the target
is to isolate it, so that its backaction noise
has no effect on $A_b$.
This is common at microwave frequencies,
where an amplifier is usually isolated behind a
circulator, with its backaction absorbed by a cold
load~\cite{Chapman_2017,Clerk_2010,Boutan:2018uoc,Brubaker:2018ebj}.
The effect on the target is as if we simply connected
it to the cold load.

In contrast to the op-amp case, where the coupling
between the amplifier and the target is taken
to be very weak, 
achieving good SNR in isolated configurations
requires stronger couplings to the target,
which significantly affect its damping properties~\cite{Clerk_2010}.
To find a SNR limit, 
we can adopt a formal perspective in which 
we view all of the degrees of freedom
apart from $A_b$ as making up the `amplifier',
and write 
$H_{\rm int} = A_b \hat F$.
Then, if the target is in its ground state
(i.e.\ its temperature and that of the cold load are
negligible),
the fluctuation spectrum of $\hat F$ must 
be such that $S_{A_b A_b} = 2 \tchi_i \Theta(\omega)$, so
$S_{FF} = 2 \frac{\tchi_i}{|\tchi|^2} \Theta(\omega)$.
Using the analysis from~\cite{Clerk_2010}, this implies
that the total noise at the amplifier
output is 
$\bar S_{A_b A_b}^{\rm tot} \ge \frac{|\tchi|^2}{2 \tchi_i}$.\footnote{
For an ideal quantum amplifier,
$\bar S^I_{A_b A_b} \bar S_{FF} \ge 1/4$.
However, in our case $S_{FF} = 0$ for $\omega < 0$,
whereas $S_{II}$ is even, which implies that there
must be wasted information in the $FI$ correlation~\cite{Clerk_2010}. In particular, by feeding back some of the output,
the back-action could be made symmetric, decreasing
$\bar S_{FF}$ by a factor $1/2$.
Consequently, in our case, $\bar S_{A_b A_b}^I \bar S_{FF} \ge 
1/2$. Thus, $\bar S^I_{A_b A_b} \ge \frac{1}{2 \bar S_{FF}}
= \frac{|\tchi|^2}{2 \tchi_i}$.
}
Consequently,
\begin{equation}
	{\rm SNR}^2 \le (g B_0 V_b)^4 t \intoinf \frac{d\omega}{2 \pi} \left(2 S_{\dot a \dot a} \tchi_i \right)^2
	\label{eqsnriso}
\end{equation}
in agreement with the limits of the circulator-plus-cold-load
analyses from~\cite{sclong,srf}.
If we assume $\tchi_i \ge M \omega \gamma |\tchi|^2$,
as discussed in the previous subsection,
then $\tchi_i^2 \le |\chi|^2 \le \tchi_i/(M \omega \gamma)$,
so we have
\begin{equation}
	\int_{\Delta m} d\omega \, \tchi_i^2 \le \frac{\pi}{2 M^2 \omega^2 \gamma}
\end{equation}
Consequently, the axion-mass-averaged ${\rm SNR}^2$ is bounded
by 
\begin{equation}
	\overline{{\rm SNR}^2} \lesssim
	\pi^2 (g^2 B_0^2 V_b \rho_a)^2 \frac{t Q_a}{m^3 \Delta m \gamma}
	\label{eqsnr2isoav}
\end{equation}
This has the same parametric form as equation~\ref{eqsnr22}.
For a single-pole response function,
we have
\begin{equation}
	\overline{ {\rm SNR}^2} \lesssim
	\frac{\pi}{4} (g^2 B_0^2 V_b \rho_a)^2 \frac{t Q_a}{m^3 \Delta m \gamma}
	\label{eqsnr2isoav2}
\end{equation}
which is $\OO(1)$ less than equation~\ref{eqsnr2isoav}.
As in the previous subsection, a single-pole response is optimal if
the response function is a convex combination
of single-pole response functions with damping rate
$\ge \gamma$, and we conjecture that it is also
optimal in the more general case of $\tchi_i \ge M \omega \gamma
|\tchi|^2$.

If the integration time becomes long compared
to the inverse bandwidth of spectral features, 
$Q \sim \nu t_1$, then we again obtain ${\rm SNR}
\sim \bar N_{\rm max}$. An isolated amplifier
satisfies the PQL assumptions, so this bound necessarily
holds.

It should be emphasised that the $\gamma$ in 
equations~\ref{eqsnr2isoav} and~\ref{eqsnr2isoav2}
refers to the damping rate with the readout system
connected to the target, and not just 
to the intrinsic damping $\gamma_{\rm int}$ in the target itself
(whereas $\gamma = \gamma_{\rm int}$ in the op-amp case).
Analysing the case of a circulator plus a cold load,
as in~\cite{sclong,srf}, shows that 
for a given intrinsic dissipation rate $\gamma_{\rm int}$
associated with the target, 
the axion-mass-averaged ${\rm SNR}$ squared
is optimised for $\gamma_{\rm det} = 2 \gamma_{\rm int}$,
where $\gamma = \gamma_{\rm det} + \gamma_{\rm int}$
(at zero temperature, and for a single-pole response function).
This leads to an SNR limit $\OO(1)$ worse than equation~\ref{eqsnr2isoav}.

So far, we have assumed that the target 
is in the ground state. If the target and the cold
load have the same non-zero temperature, then an analogous
analysis can easily be performed, incorporating
thermal fluctuations for $A_b$.
However, if the cold load is maintained at a different
temperature from the target's environment, the situation is slightly
more complicated.
As analysed in~\cite{sclong,scshort,srf}, 
if the cold load is at a lower temperature
than the target's environment, 
and the dissipation in the target system is fixed,
it is beneficial to `overcouple' to the amplifier,
increasing the damping rate of the target excitation.
If all temperatures were negligibly small, this would result in worse sensitivity,
but because it also reduces the thermal noise
reaching the detector, its overall effect is to improve
the SNR. From \cite{sclong,scshort,srf},
for $n_T \gg 1$, where $T$ is the temperature of
the target, 
the resulting ${\rm SNR}^2$ is suppressed by
$\sim 1/n_T$ compared to the $T=0$ value.

It should be possible to give a more detailed
analysis of isolated amplifier setups than
attempted in this section. Our main aim here
was to show that the SNR obtained from a 
single-pole resonator is $\OO(1)$ optimal
(at least at zero temperature).
 
\subsection{Quantum-limited op-amp}
\label{secQL1}

As discussed in appendix~\ref{seclinamp}, if it is possible to
optimise the correlations between a linear amplifier's
backaction and imprecision noise, then
the minimum added noise, referred back to the measured
variable $x$, is $S_{xx}^{\rm add}(\omega) \ge |\tchi_i(\omega)|$ (as opposed to $S^{\rm add}_{xx} \ge |\tchi|$ for uncorrelated noise).
Consequently, for an amplifier connected
in op-amp mode,
\begin{equation}
	{\rm SNR}^2 \le (g B_0 V_b)^4 t \intoinf 
	\frac{d\omega}{2\pi} \left(\frac{S_{\dot a \dot a} |\tilde \chi|^2}{2 \tilde \chi_i +S_n}\right)^2
\end{equation}
Assuming thermal noise,
$S_n = 2 n_T \tchi_i$, and taking $\tchi_i \ge M \omega \gamma |\tchi|^2$, we have
\begin{equation}
	{\rm SNR}^2 \le \frac{1}{8 \pi} (g^2 B_0^2 V_b)^2 t \intoinf
	d\omega \left(\frac{S_{\dot a \dot a}}
	{\omega \gamma (1 + n_T)}\right)^2
\end{equation}
If we take $S_{\dot a \dot a}$ to have top-hat form,
as above, then
\begin{equation}
	{\rm SNR^2} \lesssim \frac{\pi}{8} (g^2 B_0^2 V_b \rho_a)^2 t \frac{Q_a }{m_a^3 \gamma^2 (1 + n_T)^2} 
\end{equation}
This has the same form as the SQL
expression in equation~\ref{eqsnr2bband},
when
$\Delta m \lesssim (1 + 2 n_T) \gamma$;
this is as expected,
since in both cases, the added noise
is dominated by the thermal + ZPF noise.
If $\Delta m \gtrsim (1 + 2 n_T) \gamma$,
then we can improve over the SQL limit by
$\sim \frac{\Delta m}{2 n_T \gamma} = \frac{Q}{2 n_T}
\frac{\Delta m}{m}$ (for $n_T \gg 1$). This is simply the ratio
between the sensitivity bandwidth for the SQL-limited
amplifier, $\sim (1 + 2 n_T) \gamma$, and the 
broadband ($\sim \Delta m$) sensitivity
for the QL-limited case.
On-resonance, where $\tchi$ is
purely imaginary, we achieve the QL limit by having 
uncorrelated back-action and imprecision noise, 
so we expect the QL and SQL limits to coincide 
in this case.

As the above limits show, the sensitivity limit
for a quantum-limited amplifier is set by how
well we can isolate the target system from the
dissipative environment, and so reduce $\gamma$.
There does not have to be a sensitivity/bandwidth trade-off, as occurs
in the SQL and PQL cases, since as $\tchi_i$
decreases, the ZPF noise also decreases; a setup
that saturates the QL at all frequencies is
naturally broadband. 
Near a narrow resonance, this would require
the back-action/imprecision correlations
to change very fast as a function of frequency,
but far from the resonance, the required change is slower.

Conceptually, there are various amplification
schemes that can make use of correlated back-action
and imprecision noise to attain the QL
over some frequency range, e.g.
the driven non-linear cavity method proposed in~\cite{Laflamme_2010fj}.
At higher frequencies ($\gtrsim \GHz$), this
is usually practically difficult (in particular,
it is not compatible with isolation mechanisms
such as circulators, as discussed above).
However, at lower frequencies,
SQUID amplifiers can attain near-QL performance,
in some circumstances~\cite{squidhb}. 
We discuss axion detection at low frequencies ($\ll \GHz$)
in section~\ref{secQuasiStatic}.

The fact that a QL amplifier can have better
sensitivity than the PQL limit,
in some regimes, corresponds to the amplifier
enhancing the frequency-averaged quantum fluctuations of $A_b$
(similarly to how
backaction evasion effectively
drives an oscillator
into a squeezed state~\cite{Clerk_2008}). 
For a single-pole response function,
the $A_b$ quantum fluctuations have the usual value
on-resonance, but are unsuppressed off-resonance,
corresponding to the broadband sensitivity
described above.

\subsection{Up-conversion}
\label{secUpLim}

We can generalise the power absorption
calculations above to
a time-dependent magnetic field.
In this case,
\begin{equation}
	P_m \simeq \frac{g^2 V_b^2}{(2\pi)^2} \intinf d\omega 
	\, \omega \tilde\chi_i(\omega) (S_{\dot a \dot a} * S_{BB})_\omega
\end{equation}
\begin{equation}
	= \frac{g^2 V_b^2}{(2\pi)^2} \intinf d\omega \, S_{\dot a \dot a}(\omega) ((\omega \tilde \chi_i) * S_{BB})_\omega
	\label{eqsaaup}
\end{equation}
where $S_{BB}$ is the spectral density of $B_0(t)$.
Using equation~\ref{eqsintinf},
\begin{equation}
	\int dm \, P_m \simeq 
	\frac{g^2 V_b^2 \rho_a}{4 \pi } \intinf d\omega
((\omega \tilde \chi_i) * S_{BB})_\omega
\end{equation}
\begin{equation}
	\simeq g^2 V_b \rho_a \overline{B_0^2} \frac{\pi}{2}
	\label{eqsbbpow}
\end{equation}
where $\intinf d\omega \, S_{BB}(\omega) = 2\pi \overline{B_0^2}$,
so $\sqrt{\overline{B_0^2}}$ is the RMS $B_0$ value.
In the case of a static $B_0$ field, this reproduces
equation~\ref{eqpstatic}. 

There are a number of qualitatively distinct
cases, depending on the oscillation frequencies of the magnetic
field and the axion signal.
For a $B_0$ oscillation
at frequency $\omega_B$, an axion oscillation at $\omega_a$
will give a forcing at sum and difference frequencies, $\omega_B + \omega_a$ and
$|\omega_B - \omega_a|$.
To start with, we will consider the `up-conversion' case where
$\omega_B \gg \omega_a$, so that both sum and difference
frequencies are $\gg \omega_a$.

The power absorbed from single-frequency axion and
$B_0$ oscillations is set by
\begin{equation}
	P \simeq \frac{1}{4} g^2 V_b^2 B_0^2 \rho_a \omega_B
	\left(\tilde \chi_i(\omega_B + m_a) +\tilde \chi_i(\omega_B - m_a)\right)
\end{equation}
(this is valid over times $\gg m_a^{-1}$,
so that we resolve the two different frequencies).
Averaging this over an axion mass range $\Delta m$,
we obtain
\begin{equation}
	\bar P \lesssim \frac{g^2 \rho_a V_b B_0^2}{\Delta m} \frac{\pi}{8} = \frac{g^2 \rho_a V_b \overline{B_0^2}}{\Delta m} \frac{\pi}{4}
	\label{eqpup}
\end{equation}
which can be saturated by concentrating $\chi_i$
into a $\sim \Delta m$ range
either above or below $\omega_B$ (or in both ranges).
This only represents half of the total mass-integrated
power absorbed. Since axions at $m_a \simeq 2 \omega_B$
will also excite a target mode at $\simeq \omega_B$, the
other half of the absorption is at these, much higher, axion masses.

The PQL detectability
is again
set by the expected number of quanta absorbed, which is
$N_a \simeq P t / \omega_B$.
Compared to a static-field experiment with the same
$P$, an up-conversion experiment absorbs
fewer but higher-energy quanta.\footnote{
In the above, we have considered the power
absorbed from the axion-induced effective current,
since this is the quantity that is usually relevant
for detection. For a static background
magnetic field, this is the same as the power 
absorbed from the axion DM field. However,
for an up-conversion experiment,
each axion quantum absorbed results in a photon
at frequency $\sim \omega_B \gg \omega_a$; the extra
energy comes from the oscillating background magnetic 
field. 
} The SQL limit for a single-pole response (which is parametrically the same as the isolated-amplifier limit) is 
\begin{equation}
	\overline{{\rm SNR}^2} \le \frac{1}{\pi} \bar P_{\rm max}^2 
	\frac{\Delta m}{m} \frac{t Q_a Q_1}{\omega_1^3 f(n_T(\omega_1))}
	\simeq 
	\frac{1}{\pi}
	\bar P_{\rm max}^2 
	\frac{\Delta m}{m} \frac{t Q_a Q_1}{T \omega_1^2}
	\label{eqSNR2up}
\end{equation}
where $Q_1 \equiv \omega_1/\gamma$, and the second equality is for $T \gg \omega_1$.
For $n_T \ll 1$, taking $Q$ as large as
it can be while still resolving 
the resonator
bandwidth gives $\overline{{\rm SNR}^2} \sim 
\left(\frac{\bar P t_{\rm tot}}{\omega_1}\right)^2 
\sim \bar N^2$, in analogy to equation~\ref{eqSQLn1}.
A QL-limited amplifier could improve on the SQL
by the usual $\sim \frac{\Delta m}{2 n_T \gamma}$ factor. However, as we will discuss
in section~\ref{secUpExp}, up-conversion
experiments are most interesting at
$\omega_1 \gtrsim \GHz$, and utilising
backaction/imprecision correlations is difficult there.

As the limits above indicate, up-conversion experiments
 have reduced sensitivity, compared to ideal
 static-field experiments with the same $\overline{U}_B$.
However, as we will discuss in section~\ref{secQuasiStatic},
for low enough axion masses, it is hard for
static-field experiments to attain the power and sensitivity
bounds from above, due to being in the quasi-static
regime.
Consequently, up-conversion experiments can have
a parametric advantage in absorbed power,
and sensitivity, for low axion masses.

\subsection{Down-conversion}
\label{secDownLim}

We can also consider other frequency combinations.
If $\omega_B \ll m_a$, then the situation
is basically the same as the static-field case.
The remaining distinct case is when $\omega_B \simeq m_a$, so
that the difference frequency is small,
$\omega_s = |\omega_B - m_a| \ll
m_a$. 

In this case, which we label `down-conversion', 
it is possible to attain the $\bar P$ bound from
equation~\ref{eqsbbpow}, by concentrating $\tilde \chi_i$
at low frequencies $\sim \omega_s$. Then, $(\omega
\tilde \chi_i) * S_{BB}$ is entirely at frequencies $\sim \omega_B$. Thus, we obtain the same converted power
limit as for a static-field experiment.

However, since $N_a = P t / \omega_s$, by
taking $\omega_s$ smaller and smaller, we can
make the PQL sensitivity better
and better (for integration times long enough to resolve
$\omega_s$). Physically, this is the converse
of the up-conversion case --- we maintain the same
converted power, but this corresponds
to absorbing more
low-energy quanta.\footnote{From the point of view of the axion field, the $A \cdot B$ term
it is interacting with oscillates at a frequency $\sim m_a$,
whereas the effective current $g \dot a B$ oscillates at $\sim \omega_s$
--- the energy change of the axion field is therefore greater than
the energy gain in the target mode, with the difference
made up by the $B_0$ oscillation.}
As per equation~\ref{eqSNR2up}, the SQL-limited
SNR is $\propto 1/\omega_s$, for given $U_B$,
$Q$, and 
$T \gtrsim \omega_s$
(and $\propto \omega_s^{-3/2}$ for $T \lesssim \omega_s$).

Thus, at least in principle, we can improve
our sensitivity by making $\omega_s \ll m_a$.
However, as we discuss in section~\ref{secDownExp}, there
are usually serious practical obstacles
to obtaining an advantage in this way.


\section{DM detection experiments}
\label{secdmdetect}

In the previous section, 
we analysed limits
on the power absorption and
sensitivity of idealised axion DM detection experiments,
in terms of the magnetic field maintained in
the experiment.
In this section, we will investigate the sensitivities
of more specific experimental setups,
and how they relate to these theoretical limits.

\subsection{Static background field: power absorption}
\label{secStaticPower}

From equation~\ref{eqpbar1}, a static-field experiment 
covering an axion mass range $\Delta m$
must have
\begin{equation}
	\min_{m \in \Delta m} P \le 
	\pi
	\frac{g_{a \gamma \gamma}^2 \rho_a \overline{U}_B}{\Delta m}
	\label{eqsb1}
\end{equation}
if it is in the linear response regime, and
if $\tchi_i (\omega) \ge 0$ for $\omega > 0$.\footnote{More
generally, eq.~\ref{eqsb1} applies if $\tchi_i$ is well-approximated,
in the frequency range $\Delta m$,
by the response function $\tchi_i'$ for a system
which has $\tchi_i' \ge 0$ for $\omega \ge 0$.}
For experiments with $m_a \gtrsim L^{-1}$,
where $L$ is the length scale of the shielded volume,
it is easy to see that the usual EM field modes
$\OO(1)$-saturate the sum rules, for slowly-varying
$B_0$ profiles. Consequently, the signal power
for cavity and dielectric haloscopes is parametrically
given by equation~\ref{eqsb1}.
For example, a cavity haloscope has an on-resonance,
fully-rung-up signal power of
\begin{equation}
	P_{\rm sig} = C g^2 B_0^2 V Q \frac{\rho}{m}
\end{equation}
where $Q$ is the quality factor of the cavity mode, and
\begin{equation}
	C \equiv \frac{\left|\int dV \, E \cdot B_0\right|^2}{
		\int dV |B_0|^2
		\int dV |E|^2}
\end{equation}
is the normalised overlap between the background magnetic field
and the electric field of the mode.
Geometrically, $C \le 1$.
If $Q \gg 1$, then the converted power is approximately a
Lorentzian function of the axion mass, so if we want to cover
a small mass range $\Delta m$ with cavity configurations
tuned to different resonant frequencies, we have
\begin{equation}
	\bar P \simeq \frac{g^2 B_0^2 V_b \rho}{\Delta m} \frac{\pi}{2} C
\end{equation}
as expected from equation~\ref{eqsb1}.
ADMX~\cite{Du:2018uak,Braine:2019fqb}
uses the ${\rm TM}_{010}$ mode of a cylindrical cavity,
which has $C \simeq 0.68$ (ignoring perturbations
from the tuning rods),
so is $\OO(1)$ optimal for its cavity volume.
The rest of the sum-rule-determined absorption
is into other cavity modes --- these will generally be
at different frequencies, so will not be useful for searching
a small axion mass range (though they may be used to perform
simultaneous searches in different mass ranges,
e.g.\ ADMX's usage of the ${\rm TM}_{020}$ mode alongside
${\rm TM}_{010}$~\cite{Boutan:2018uoc}).

For a dielectric haloscope with alternating half-wave layers
of refractive indices $n_1, n_2$, 
the average signal power over an axion mass range
$\Delta m$ around the half-wave frequency is~\cite{Baryakhtar:2018doz}
\begin{equation}
\bar P = \frac{4}{\pi} \frac{g^2 B_0^2 V \rho}{\Delta m}
	\left(\frac{1}{n_2} - \frac{1}{n_1}\right)^2
	\label{eqPlayers}
\end{equation}
If we take $n_1 = 1$, $n_2 \nearrow \infty$, then this is
$\sim 8/\pi^2 = 0.81$ of the sum rule limit.
The rest of the response function will be at other axion
masses. For a half-wave stack, this will mostly be
at odd multiples of 
the half-wave frequency~\cite{Baryakhtar:2018doz}.
Since the the absorbed power at each multiple
drops off $\propto 1/\omega^2$, and
$1/\left(\sum_{n=0}^\infty \frac{1}{(2n+1)^2}\right) = \frac{8}{\pi^2}$, this agrees with the $8/\pi^2$ fraction
absorbed into the first multiple 
that we found from equation~\ref{eqPlayers}.

At microwave frequencies, it is relatively simple
to find low-loss transparent dielectrics, which can almost
saturate these bounds --- for example, the MADMAX~\cite{TheMADMAXWorkingGroup:2016hpc,Brun:2019lyf} proposal currently aims to
use LaAlO$_3$ layers, with permittivity $\epsilon \simeq 24$.
However, at higher frequencies, it
may be difficult to find suitable dielectrics
with large $n$. For example, in the energy range
$\sim 0.2 - 0.5 \eV$,
a potentially practical pair of transparent
dielectrics is Si ($n \simeq 3.4$) and ${\rm Al}_2 {\rm O}_3$
($n \simeq 1.77$). 
This gives a suppression of
$\frac{8}{\pi^2}
	\left(\frac{1}{n_2} - \frac{1}{n_1}\right)^2 \simeq 0.07$
for the absorbed power, relative to the volume-limited value.

In addition, for a small number of layers, the shielded
volume required may be significantly larger than the
dielectric-occupied volume, e.g.\
to allow space for the focussing optics. In that
sense, the experiment is even further from optimality.\footnote{Techniques such as using
microlens arrays rather than a single lens~\cite{ChilesPersonal} may help
with volume utilisation.}
However, such experiments can still be efficient
compared to other high-frequency proposals, such as dish
antennae~\cite{Horns:2012jf}.
For example, if we consider the Si/Si${\rm O}_2$ chirped
stacks with reach shown in figure~\ref{figdpsens} (each
of these has 100 periods, covers an octave in axion mass
range, and has area $\sim (18 \cm)^2$),
an equivalent reach could be obtained from
a $\sim {\rm meter}^2$ dish antenna in the same background
magnetic field.
This dish would require a significantly larger
volume for the magnetic field, and could only concentrate
the emitted photons onto a significantly larger area, complicating
detection. If larger volumes of photonic material could be fabricated,
or larger refractive index contrasts achieved, the comparison
would be even more favourable to the layered materials.

The cavity and dielectric haloscope examples
above achieved $\bar P$ values at $\OO(1)$
of the limit. An example of a target that could theoretically
attain the limit almost fully (for a uniform
background magnetic field, within a narrow axion mass range)
is a uniform plasma, of large extent compared 
to the axion Compton wavelength (as analysed
in~\cite{Lawson:2019brd}). It should be noted
that the wire meta-materials proposed in~\cite{Lawson:2019brd}
most likely do not have this property, and
have power absorption at $\OO(1)$ of the limit,
similarly to a cavity haloscope.

The sum rule limits
explain some basic properties of axion conversion
rates in experiments; in particular, 
the power/bandwidth tradeoff,
and the similar axion-mass-averaged signal
powers attained from different types of experiments
with comparable volumes and magnetic fields.
They also imply less obvious facts.
For example, in dielectric haloscopes, one can enhance
the conversion rate near a particular frequency
by making the dielectric permittivity $\epsilon$ of some layers $\epsilon < 1$,
e.g.\ by operating near an optical resonance (the stack configuration of the molecular
absorption proposal in~\cite{Arvanitaki:2017nhi} is an example of this). However, the sum rules
tell us that this is compensated for by lower powers at other frequencies,
and cannot improve the average power over an $\OO(1)$ axion mass range.

As mentioned in section~\ref{secSummary},
static-field experiments for $m_a \ll L^{-1}$ generally
suffer a quasi-static suppression,
and the maximum power absorbed is suppressed by
$\sim (m_a L)^2$. We defer discussion of this regime to
section~\ref{secQuasiStatic}.

\subsection{Detectability}

\subsubsection{Linear amplifiers}

Section~\ref{secTheorySens} showed that,
for systems satisfying the PQL assumptions,
or for readout via a SQL or isolated amplifier, 
the theoretical detectability limits are closely
related to the limit on $\bar P$.
With SQL op-amp readout, it is easy to imagine
idealized setups that attain the SNR bound from
equation~\ref{eqsnr22sp}, to $\OO(1)$. 
For example, in a cavity
haloscope experiment, we could connect an
antenna to a SQL-limited current amplifier. 
In the case of a linear amplifier isolated 
behind a circulator and cold load, 
it is also easy, in principle,
to attain the bounds from section~\ref{secIsolated}
to $\OO(1)$~\cite{srf}.

There have been claims in the literature that
simple feedback schemes can give enhanced sensitivities;
for example, the digital feedback scheme
of~\cite{Daw:2018qwb}.
Since the feedback applied in~\cite{Daw:2018qwb} is entirely coherent-state,
and a SQL (or circulator-isolated) readout system
seems to be assumed, the bounds derived above
should apply. In particular, the 
thermal-noise-limited sensitivity cannot
be improved by adding a known, coherent signal
to the mode.
Moreover, attempting to cover
multiple frequency ranges in parallel results
in worse theoretical sensitivity than
tuning a narrow resonance over time,
as per the results of section~\ref{secTheorySens}.

\subsubsection{PQL}
\label{secPQL}

In the PQL case, it is also simple to describe setups
that can (in principle) attain
the $N_a = \bar P t / m_a$ limit;
for example, 
if we have no added noise ($T \ll m_a$ etc), and almost
all of the converted power is absorbed by a background-free
photon counter. 

One motivation for discussing the PQL, as opposed to
specialising to e.g.\ absorptive photon counters
directly, is that it can also apply to experiments
using different kinds of target excitations, such
as phonons~\cite{Knapen:2017ekk}, electron-hole
pairs~\cite{Agnese:2018col}, or more complicated
quasi-particles~\cite{Marsh:2018dlj}
(and different detection schemes
such as optical homodyne detection). 
While a detection setup always includes
some components that are not in thermal equilibrium 
(e.g.\ amplifiers),
in many cases, these are connected to the
target system via a damping-type coupling. Examples
include absorptive photodetectors, or sensors
isolated behind a circulator with a matched load.
In these cases, the quantum fluctuations of the target
system are as if the sensor were replaced by an equivalent
passive load. Consequently, the experiment's
sensitivity should be bounded by the PQL
(for example, this applies directly to the
case of an SQL amplifier isolated behind a
circulator, as discussed above).

As mentioned in section~\ref{secSumDet}, it is 
possible to violate the PQL by
preparing the EM field in a `non-classical' state, e.g.\
a Fock state or a squeezed state. It is also the case that
some detection schemes push the target into an PQL-violating
state; examples include back-action evasion~\cite{Clerk_2008}, QND
photon counting, and (as per the QL discussions above)
correlated backaction/imprecision noise.

\subsection{Quasi-static regime}
\label{secQuasiStatic}

When $m_a \ll L^{-1}$, it is significantly more
difficult to attain the sum-rule bounds
with a static-field experiment. This is because
the EM fields 
are naturally in the quasi-static
regime at frequencies $\ll L^{-1}$,
and their $A$ fluctuations are suppressed.
The magnitude of the linear response function is
similarly suppressed, so the amplifier SNR is also
affected.

We can demonstrate this suppression, somewhat heuristically,
by considering the fluctuations of the energy in the
electromagnetic field.
The rate of change of EM field energy is
\begin{equation}
	P_{\rm EM} = - \int dV \, E \cdot J
\end{equation}
In Lorenz gauge,
$\nabla \cdot A = - \partial_t A_0$,
we have $\partial_\mu \partial^\mu A^\nu
= (\partial_t^2 - \nabla^2) A^\nu = J^\nu$.
If we are considering very low
frequencies, $\omega \ll L$, then since the $A^\nu$ and $J^\nu$ fields
are localised on scales $\sim L$, we have $\nabla^2 A^\nu \simeq - J^\nu$.
Hence,
\begin{equation}
	P_{\rm EM}  \simeq \int dV \, E \cdot (\nabla^2 A)
	\label{eqpem}
\end{equation}

We can ignore the low-frequency condition for a moment,
and take the expectation value of equation \ref{eqpem}'s RHS,
in the ground state.
As per section~\ref{secAxVel}, the terms corresponding
to spatially orthogonal $A$ profiles add independently,
and have the same sign.
Since
\begin{equation}
	- \int dV b \cdot (\nabla^2 b) =
	\int \frac{d^3k}{(2 \pi)^3} k^2 |b_k|^2
	\gtrsim L^{-2} V_b
\end{equation}
we have
\begin{equation}
	-i \int dV \langle E_b \cdot (\nabla^2 A_b) \rangle
	\gtrsim - i  \langle E_b A_b \rangle \frac{V_b}{L^2}
	= \frac{1}{2 L^2}
\end{equation}
Similarly, if we consider the fluctuations within a particular
frequency range, we have
\begin{equation}
	\Delta P_{\rm EM} \gtrsim \frac{V_b}{L^2} \frac{1}{2 \pi}
	\int_{\Delta \omega} d\omega \, \omega \, S_{A_b A_b}(\omega)
\end{equation}
Hence, if the sum rule were $\OO(1)$-saturated
by the low-frequency fluctuations, then
the low-frequency fluctuations of $P_{\rm EM}$ would
be $\gtrsim \frac{1}{2 L^2}$, in the ground state. In contrast,
if the low-frequency EM modes
behave like harmonic oscillators, then $P_{\rm EM} \sim \omega^2$.
Hence, in the latter case, the average value of $\SC_{A_b A_b}$
must be suppressed by $\sim (\omega L)^2$ from its
sum-rule-limited value.

We can gain some physical intuition for this suppression by
splitting the quasi-static electric field into 
a `Coulomb' part $E_C = - \nabla A_0$,
and a `Faraday' part $E_F = - \partial_t A$ (again, in Lorenz gauge).
The Coulomb part can have large fluctuations (e.g.\
for oscillations involving a capacitor),
but $E_C$ has zero integrated overlap with $B_0$.
Conversely,
$E_F$ is naturally $\sim (\omega L) B$, where $B$ is the magnetic
field fluctuation. Hence, if the ground state
magnetic field fluctuations in a mode
carry $\sim$ one quantum of energy, which is
the case for harmonic-oscillator-like modes,
then the $E_F$ fluctuations will be suppressed,
as will the $A$ fluctuations.

Given this suppression, a static-field experiment searching
for an axion over a small mass range $\Delta m$ around $m_a$ must have
\begin{equation}
	\min P \lesssim \frac{g^2 B_0^2 V_b \rho_a}{\Delta m}
	(m_a L)^2
\end{equation}
This scaling can be confirmed by calculations
for specific experimental setups.
In \cite{Chaudhuri:2014dla}, the response of a small cavity in
a constant magnetic field was computed, and 
in~\cite{Kahn:2016aff} a toroidal magnetic field was considered,
both displaying the expected $\sim (m L)^2$ suppression.

The suppressed power absorption directly limits the sensitivity
of PQL-limited searches.
For SQL searches, the $\sim (m L)^2$ suppression
of the frequency-averaged $\tchi_i$ value
contributes a $\sim (m L)^2$ suppression in ${\rm SNR^2}$,
while another factor comes from
$U_E/U \sim (m L)^{-2}$ in equation~\ref{equue}, since
the electric field fluctuations are only a 
$\sim (m L)^2$ fraction of the total energy of 
a $\omega \simeq m$ oscillation. 
For QL searches, the $U_E/U$ suppression
itself suppresses the ${\rm SNR}^2$ by $\sim (m L)^4$.
Consequently, for all of these types of searches,
the $g$ sensitivity is suppressed by $\sim (m L)^{-1}$
compared to its higher-frequency scaling.\footnote{The
proposal in \cite{McAllister:2018ndu} aims to avoid
this suppression using a simple capacitive pickup.
This relies on an alternative interpretation of
axion electrodynamics, in which the axion
is not necessarily derivatively coupled to SM
fields~\cite{Tobar_2019,Tobar_2019_2}.}

\cite{sclong,scshort} conducted a detailed
analysis of SQL-limited detection in the quasi-static
regime, assuming a given axion-induced flux coupled
through a pickup loop of given inductance.
In agreement with our analysis, the ${\rm SNR}^2$ limit 
they obtain is
parametrically given by equation~\ref{eqsnr22sp},
suppressed
by $(m L)^4$. 

As well as the resonant approaches of~\cite{sclong,scshort},
broadband approaches to low-frequency axion DM
detection have been proposed, such as ABRACADABRA~\cite{Kahn:2016aff}. This intends to use SQUID amplifiers,
which can,
in some circumstances,
achieve near-QL
sensitivity~\cite{squidhb}. If a QL-limited broadband experiment
could be realised, it would have superior
sensitivity to a SQL-limited resonant search, as discussed
in section~\ref{secQL1} (the comparison of resonant to broadband
approaches in~\cite{sclong,scshort} was based on both being
SQL-limited).
In practice, achieving the required amplifier
properties would likely be very difficult 
(for a large pickup loop, with a correspondingly large
inductance, an amplifier with a matching noise
impedance would be required).
Figure~\ref{figAxionLim} compares
the PQL, SQL and QL sensitivites in the quasi-static regime,
illustrating these differences.

\subsubsection{Evading quasi-static limits}

It should be noted that, to escape the `quasi-static' regime
where $\nabla^2 A \simeq -J$, not all of the dimensions need to be large. For example, the
short cylinders proposed in~\cite{Daw:2018qwb} can attain the sum-rule
bound to $\OO(1)$, as could a one-dimensional transmission line with
length $\gtrsim m_a^{-1}$.

Even for an experiment with all dimensions $\ll m_a^{-1}$, 
it is still possible to attain the sum-rule bound if we can enhance
the EM energy fluctuations.\footnote{
	This shows that one has to be careful
	in interpreting the results of papers such
	as \cite{Ouellet:2018nfr,Beutter:2018xfx},
	which show that the `induced EM fields' from the
	axion DM oscillation are suppressed, for 
	background magnetic fields of small extent
	compared to the axion Compton wavelength.
	This is true, as defined, but does not have to lead to
	a smaller converted power, or worse detectability.
For example, \cite{Ouellet:2018nfr} makes the assumption that
\emph{``Our detector is composed of a collection of time-independent charges and currents, $\rho_e$ and $J_e$''}, which automatically excludes
e.g.\ resonant receivers.} Conceptually, the simplest
way to accomplish this would be to use a material
with high magnetic permeability $\mu$~\cite{sclong}. Then, a magnetic
field $B$ has EM energy density $B^2/2$, but the magnetisation
$M = (1-\mu^{-1}) B$ contributes a negative energy density
$-(1-\mu^{-1})B^2/2$, giving total energy density $B^2/(2 \mu)$. If the
total magnetic energy (including the magnetisation term) of the mode has
fluctuations $\sim \omega$, then the EM field contribution has fluctuations
$\sim \mu \omega$, `borrowing' energy from the spins. If we make $\mu
\sim 1/(\omega L)^2$, we can attain the sum-rule bound (at larger $\mu$,
the wavelength $\lambda \sim \frac{2 \pi}{m \sqrt{\mu}}$
of the EM modes becomes comparable to $L$~\cite{sclong}).

Circuit elements which can draw energy from
e.g.\ a DC bias current, or a magnetic field bias,
can also provide energy for the
target mode's magnetic field to `borrow' from,
enhancing its fluctuations. A toy example would
be to connect an element with a negative inductance,
such as a flux-biased loop~\cite{Semenov_2003},
in series 
with the physical pickup inductor. In that case, the
energy in the loop fluctuates downwards, as the physical inductor's magnetic field
energy fluctuates upwards.

If such a setup has $\tchi_i \ge 0$ for $\omega > 0$,
which must be true if it cannot on average emit
power into the axion field,
then it cannot beat the usual 
$\bar P$ bound of equation~\ref{eqsb1}.
For example, if we tried to use a negative
inductor to cancel the inductance of the physical
pickup loop to even higher precision than required
to attain the $\bar P$ bound, then the energy
of the $E_F$ field due to the changing magnetic field
through the pickup loop would become important.
The pickup loop necessarily stops behaving like
an ideal inductance at some point, in analogy to the 
wavelength limit
on enhancement from permeable materials.
Similarly, if $\tchi_i$ is well-approximated, in
the relevant frequency range, by 
a response function $\tchi_i'$ that has
$\tchi_i' \ge 0$ for $\omega > 0$, then 
the power absorbed is also bounded by equation~\ref{eqsb1}.

The methods discussed above are interesting in principle,
but would likely encounter practical difficulties.
High-permeability
materials, in particular, have major noise issues; they are generally lossy, have significant hysteresis and
spin noise, and behave pathologically at high
($\gtrsim \MHz$) frequencies~\cite{sclong,SaptarshiPersonal}.
Approaches such as circuits with negative differential
resistances (e.g.\ negative inductors) seem more promising,
but may still have noise issues --- for example, magnetic
flux noise through flux-biased loops~\cite{SaptarshiPersonal}. 
Whether setups with $\omega \tchi_i \ge 0$,
which are limited by sum rules, would be easier
to implement than more general active approaches,
is another interesting question.
As discussed in the next subsection, 
a different approach to avoiding the quasi-static suppression
at low axion masses is via up-conversion.

\subsection{Up-conversion}
\label{secUpExp}

Up-conversion experiments have been proposed at both microwave~\cite{Sikivie:2010fa,Goryachev:2018vjt} and optical~\cite{DeRocco:2018jwe,Obata:2018vvr,Liu:2018icu} frequencies.
In the case of microwave cavities, it is simple to see that they can
attain the sum rule bounds from section~\ref{secUpLim}, to $\OO(1)$.
Suppose that we have a narrow-bandwidth axion signal,
$a(t) = a_0 \cos (m_a t)$.
By equating $P_{\rm in} = -\int dV E \cdot J^{(a)}$ 
to $P_{\rm loss} = \frac{U \omega}{Q}$, we can obtain
the formula from~\cite{Sikivie:2010fa},
\begin{equation}
	P_{\rm sig} = \frac{1}{8} (g a_0)^2 m_a^2 \frac{Q_1}{\omega_1} 
	\frac{\left(\int dV E_1 \cdot B_0 \right)^2}{\int dV E_1^2}
\end{equation}
for the fully-rung-up, on-resonance signal power. Here,
$B_0$ is the amplitude of the oscillating magnetic field, $B(t) = B_0 \cos (\omega_B t)$.
This gives a mass-averaged power of
\begin{equation}
	\bar P \simeq \frac{\pi}{8} g^2 B_0^2 \frac{\rho}{\Delta m} 
	\frac{\left(\int dV E_1 \cdot b_0 \right)^2}{\int dV E_1^2}
	\equiv \frac{\pi}{8} g^2 B_0^2 \frac{\rho}{\Delta m} 
	C_{01} V_{b_0}
\end{equation}
Since $C_{01} \le 1$, with equality iff $E_1$ and $B_0$
have the same profile, this agrees with equation~\ref{eqpup}.
For example, the small-scale up-conversion experiment we
discuss in~\cite{srf}, which drives the 
${\rm TE}_{012}$ mode of a cylindrical cavity, and picks
up signals in the 
${\rm TM}_{013}$ mode, has $C_{01} \simeq 0.19$.
The optical interferometry experiments proposed in \cite{DeRocco:2018jwe,Obata:2018vvr,Liu:2018icu} are phrased in somewhat different terms, 
but have the same parametric behaviour, with
the orthogonally polarised drive and signal modes having $C_{01} \simeq 1$.

For $m_a \gtrsim L^{-1}$, where static-field experiments
would be outside the quasi-static regime, an up-conversion
experiment with the same volume and RMS magnetic field as a
static-field experiment can have parametrically the same
$\bar P$ value.
However, as mentioned in section~\ref{secSummary},
the achievable magnetic fields in up-conversion experiments will generally be smaller,
being limited by both material properties and cooling systems. The driving required for the oscillating magnetic
field will also introduce noise issues~\cite{srf}. Furthermore,
the cooling power required will generally restrict
the attainable physical temperature to $\gtrsim 1\kelvin$~\cite{srf}. Consequently,
static-field experiments will generally be superior.
A potential benefit of up-conversion experiments is that
the magnetic field is entirely internal to the cavity,
so superconducting cavities with very high quality factors
can be used. For a static-field experiment, using an external
magnetic field requires either a normal cavity, or a superconducting cavity
in a vortex state~\cite{Di_Gioacchino_2019}, both of which have worse quality factors.
However, this difference will generally not be enough to compensate
for the disadvantages listed above.

\begin{figure*}[t]
	\begin{center}
		\includegraphics[width=.8\linewidth]{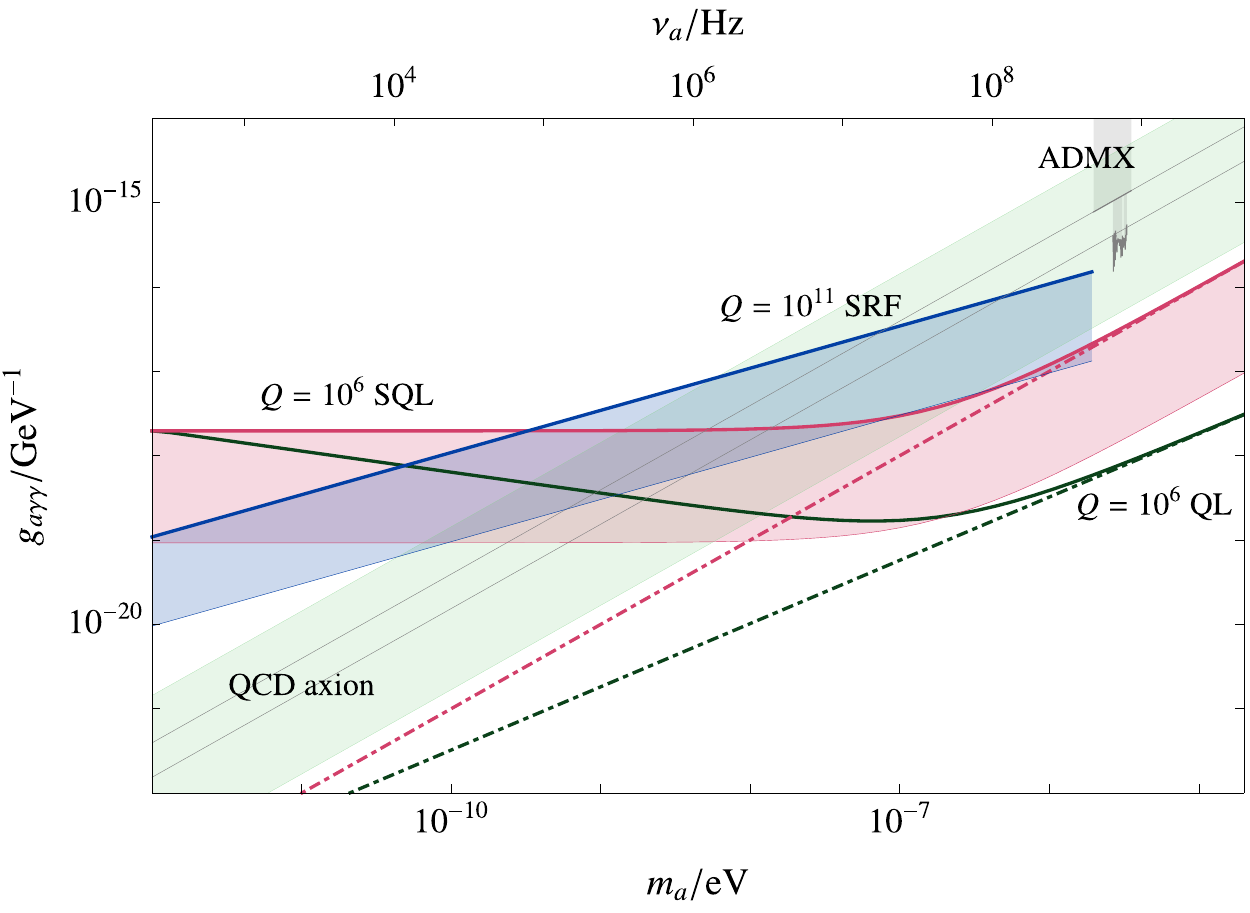}
		\caption{Theoretical sensitivity limits for experiments searching for axion DM through its coupling to photons, $\LL \supset g_{a \gamma \gamma} a
		E \cdot B$. 
		We assume that axion DM has a virialised
		velocity distribution with $\delta v \sim 10^{-3}$,
		and density at Earth of $\rho_a \simeq 0.3 \GeV \cm^{-3}$.
		The red and dark green lines correspond to experiments using
		a static $4 \Tesla$ background magnetic field, with
		an experimental volume of $1 \metre^3$, and an integration
		time of one year per $e$-fold in axion mass. The blue
		lines correspond to up-conversion experiments,
		with a target frequency of $1 \GHz$, and
		taking a RMS background magnetic
		field strength of $0.2 \Tesla$ in a volume of $1 \metre^3$
		(with the same integration time assumptions).
		The $Q = 10^6$ SQL line (solid red) assumes a physical
		temperature of $10 {\rm \, mK}$, a quality
		factor of $10^6$, and SQL-limited op-amp readout
		(we choose an expected SNR value of 3 as our sensitivity
		threshold, in this and other cases).
		The solid red line assumes
		that the experiment operates in the quasi-static
		regime, with a consequent suppression
		in sensitivity (see section~\ref{secQuasiStatic}),
		while the dot-dashed line 
		shows the full sum-rule-limited
		sensitivity (e.g.\ from using an appropriate
		matching circuit).
		The thin red line at the bottom of the red shaded
		region corresponds to the PQL-limited sensitivity
		(assuming quasi-static suppression); 
		for higher quality factors,
		SQL-limited readout can approach this.
		The $Q = 10^6$ QL line (solid green) takes
		the same assumptions as the SQL line, but with
		quantum-limited readout (see section~\ref{secQL1}),
		while the green dot-dashed line shows this without
		the quasi-static suppression.
		The $Q=10^{11}$ SRF line (solid blue)
		corresponds to the 
		isolated-linear-amplifier sensitivity for an
		up-conversion experiment with
		physical temperature $1.5 \kelvin$, and mode quality factor $10^{11}$~\cite{srf}.
		The thin blue line at the bottom of the blue
		shaded region  shows the
		PQL-limited sensitivity for these parameters 
		--- again, for higher quality factors,
		isolated amplifier readout can approach
		this~\cite{srf}.
		The green diagonal band corresponds to the `natural' range of
		$g_{a\gamma\gamma}$ values at each QCD axion mass ---
		if we write $g_{a \gamma \gamma} = \frac{\alpha_{\rm EM}}{2 \pi f_a}\left(\frac{E}{N} - 1.92\right)$~\cite{diCortona:2015ldu},
		then the upper edge of the band is at $E/N = 5$~\cite{DiLuzio:2016sbl},
		and the lower edge at $E/N = 2$~\cite{diCortona:2015ldu}.
		The gray diagonal lines indicate the KSVZ (upper, $E/N = 0$)
		and DFSZ (lower, $E/N = 8/3$) models.
		The gray shaded region corresponds to the parameter
		space excluded by ADMX~\cite{Du:2018uak,Braine:2019fqb}.}
	\label{figAxionLim}
	\end{center}
\end{figure*}

For $m_a \lesssim L^{-1}$, static-field experiments
are generically in the quasi-static regime,
as discussed above, and have coupling sensitivity suppressed
by
$\sim (m L)^{-1}$. Up-conversion experiments do not suffer
from this suppression, so can scale
better at axion frequencies $\ll \GHz$ (for lab-scale
experiments).
Figure~\ref{figAxionLim} shows a quantitative version
of this comparison. For the static-field experiments,
we take nominal parameters of $V = 1 \metre^3$,
$B_0 = 4 \Tesla$, and $T = 10 {\rm \, mK}$.
For the nominal SRF experiment,
we take $\sqrt{\overline{B_0^2}} = 0.2 \Tesla$
(at the higher end of values that might be achievable
with nioibium cavities~\cite{srf}), $V = 1 \metre^3$,
and $T = 1.5 \kelvin$ (as discussed in~\cite{srf},
cooling requirements make sub-kelvin temperature impractical).
In a realistic experiment, there would likely be
some form factor suppression (e.g.\ the $C \simeq 0.2^2$
estimate for static-field setups in~\cite{scshort}); figure~\ref{figAxionLim} displays the volume-wise
limits, both so as to set a lower bound
on the coupling sensitivity, and since our main point is to illustrate
the different scalings. All of the curves
use the single pole sensitivity limits.

As investigated in~\cite{sclong,scshort}, the SQL-limited
sensitivity for searches in the quasi-static regime is
constant as a function of axion mass. For
$m_a \gtrsim T/Q_1$, the QL-limited coupling sensitivity is
better, and scales $\propto m_a^{-1/4}$
in the quasi-static regime.
Conversely, the up-conversion coupling sensitivity
scales $\propto m_a^{1/2}$, so is theoretically
better at very low axion masses (of course,
additional noise sources, e.g.\ vibrations,
may make practical measurements difficult here~\cite{srf}).
Due to the smaller $B_0$ field, the SRF sensitivity
is significantly
worse at high ($\sim \GHz$) frequencies,
where the quasi-static suppression is not significant.
We also plot the static-field sensitivities without
the quasi-static suppression (e.g.\ given a matching
circuit of the kind discussed in the previous section),
showing that, as expected, both the scaling
and absolute sensitivity are superior
to up-conversion in these circumstances.
For measurement schemes which violate
the assumptions of the different limits, e.g.\
quantum measurement approaches such as backaction
evasion, different analyses would apply.

In our companion paper~\cite{srf}, we calculate
basic sensitivity estimates for some specific 
SRF up-conversion setups, attempting to take
into account some possible noise sources.
Comparing these to the projections from
static-field experiments, we find that,
with existing technologies, static-field experiments
(even in the quasi-static regime)
will most likely have better QCD axion reach.
At very low axion masses, where the theoretical
advantage of up-conversion experiments
is greatest, more careful investigation of noise
sources
would be required to understand whether improving
on static-field experiments is plausible.
Nevertheless, given the strong motivations
for exploring axion parameter space, and
the technological developments that may occur,
it is important to understand the properties of 
different experimental approaches.

Comparing SRF setups to optical up-conversion
experiments, the latter
suffer from two major disadvantages compared
to microwave frequencies. The first is that achievable electromagnetic
field strengths at optical frequencies are much lower.
Taking the optimistic parameters from~\cite{Liu:2018icu},
we can consider a 40m long optical cavity with circulating optical power $\sim 1 {\rm \, MW}$ (for comparison,
the circulating power in the LIGO interferometer arms is $\sim 100 {\rm \, kW}$~\cite{TheLIGOScientific:2016agk}). 
This corresponds to a stored magnetic field energy of $\sim 0.1{\rm \, J}$. 
For comparison, the small-scale SRF experiment discussed
in our companion paper~\cite{srf} has stored energy $\sim {\rm \, kJ}$,
while the nominal up-conversion experiment in
Figure~\ref{figAxionLim} has $\sim 15 {\rm \, kJ}$.
The other issue is that, for an given signal power, the number
of signal photons is much lower, by a factor
$\sim \eV / (2 \pi \GHz) \sim 2 \times 10^5$.
As a result, the theoretical sensitivity limits
for optical experiments are significantly worse,
as illustrated in Figure 2 of~\cite{srf}.
Of course, optical experiments may be cheaper or simpler
to implement than static-field or SRF experiments.

Microwave up-conversion experiments were also proposed
in~\cite{Tobar_2020,Goryachev:2018vjt}.  These have
sensitivity projections which, at low axion masses,
are significantly below our limits. As in the case of \cite{McAllister:2018ndu},
it appears that this arises from an alternative interpretation of axion
electrodynamics, in which the axion is not necessarily derivatively
coupled to SM fields~\cite{Tobar_2019,Tobar_2019_2}.

\subsection{Down-conversion}
\label{secDownExp}

As discussed in section~\ref{secDownLim}, it is theoretically possible
to increase $\bar N_a$ (and improve amplifier SNR) by
using an oscillating $B_0$ field with frequency $\simeq m_a$, and
concentrating the target mode's
fluctuations at low frequencies. However, at frequencies
$\omega_s \ll L^{-1}$, we are naturally in the quasi-static regime,
where the fluctuations are suppressed by $\sim (\omega_s L)^2$.
In that case, the down-conversion sensitivity is $\bar N_a \propto \omega_s$,
and drops at lower frequencies. Hence,
unless we circumvent the quasi-static regime,
down-conversion experiments are only of interest
for $\nu_a \gg \GHz$, where they could offer an idealised
sensitivity up to $\sim \nu_a / \GHz$ times better than static-field
experiments with the same $B_0$ amplitude.
It is easy to think of toy implementations
which can attain this bound.
For example, if $\nu_a$ were in the optical regime, we could fill
a microwave cavity with a photonic material, with periodicity
at optical wavelengths. If a laser at the frequency of periodicity were
shone through the cavity, then in the presence of an axion oscillation at almost the same frequency, the axions/laser photons would down-convert to microwave
photons.

An obvious issue with these proposals is that achievable high-frequency
($\gg \GHz$)
magnetic fields are orders of magnitude smaller than static ones, such that
even an idealised advantage
does not seem practical.
Taking the optical-frequency 40m cavity parameters from the previous 
subsection, the $\eV / (2 \pi \GHz) \sim 2 \times 10^5$
enhancement is not enough to compensate for the decreased
magnetic field energy, compared
to a nominal static-field experiment with
$U_B \sim {\rm Tesla}^2 \times \mbox{meter}^3 \sim {\rm \, MJ}$.
Consequently, even with these extreme parameters, 
no conversion rate advantage would be achieved (even
before considering the extra difficulties of detecting microwave photons
vs optical photons).


\begin{figure*}[t]
	\begin{center}
		\includegraphics[width=.8\linewidth]{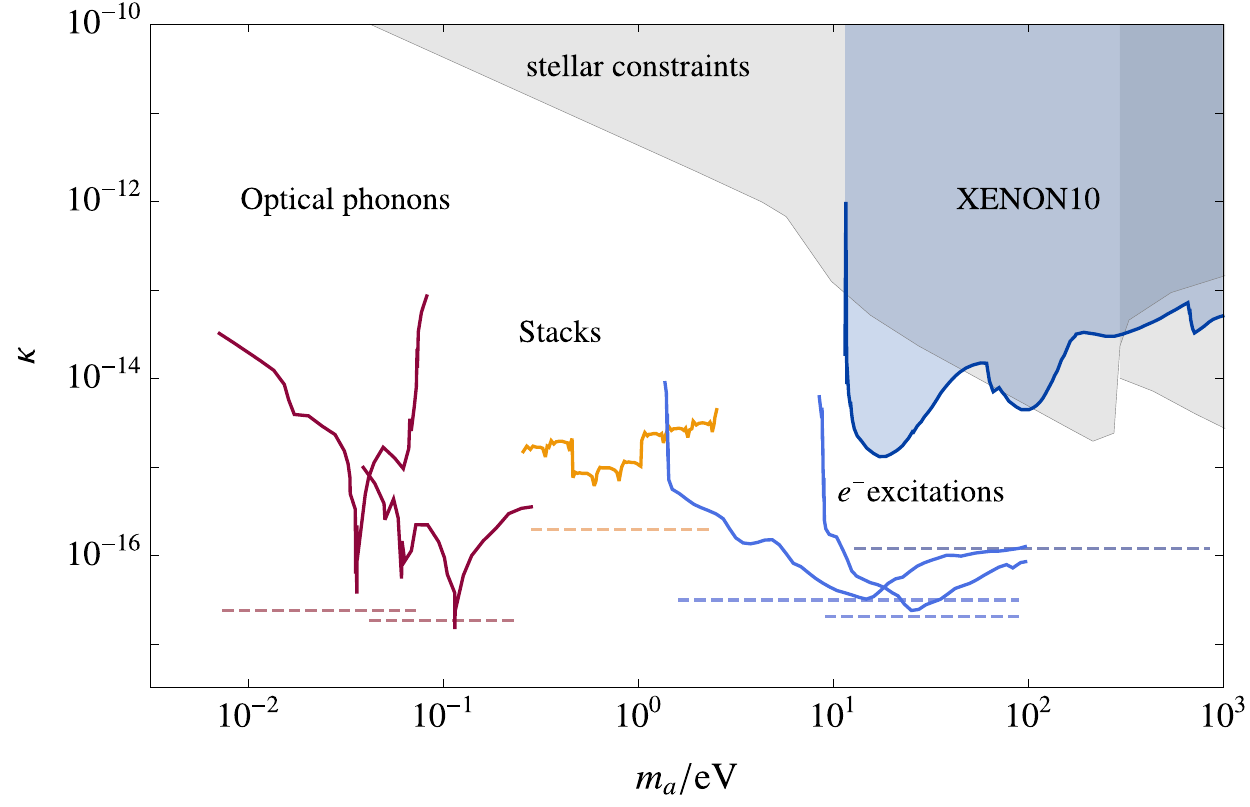}
		\caption{Sensitivity projections for
		some existing and proposed dark photon DM
		detection experiments (solid lines), along with PQL sensitivity
		limits for the corresponding target volumes and
		integration times (dashed lines). 
		See section~\ref{secDP} for descriptions of the different
		projections.
		We assume that the dark photon DM has density at Earth of $\rho \simeq 0.3 \GeV \cm^{-3}$.
		The gray `stellar constraints' region
		corresponds to the bounds on
		energy loss from the Sun and from
		horizontal branch stars~\cite{An:2013yfc,Vinyoles:2015aba,An:2013yua}.
		}
	\label{figdpsens}
	\end{center}
\end{figure*}


\section{Dark photon DM}
\label{secDP}

Apart from spin-0 candidates such as axions, light DM could
also consist of oscillations of a vector field. 
Similarly to spin-0 DM, there are a variety of possible non-thermal production
mechanisms, including purely-gravitational production
from fluctuations `stretched' by inflation~\cite{Graham:2015rva}.
The simplest, and (at low vector masses) the least constrained, coupling
of such a vector to the SM is the `kinetic mixing' coupling,
\begin{equation}
	\LL \supset -\frac{1}{4} F_{\mu\nu} F^{\mu\nu} - 
	\frac{1}{2} \kappa F_{\mu\nu} F'^{\mu\nu}
	- \frac{1}{4} F'_{\mu\nu} F'^{\mu\nu} + \frac{1}{2} m^2 A'_\mu 
	A'^\mu
\end{equation}
for a `dark photon' $A'$. 
This is equivalent, after a field 
redefinition, to a massive vector with a small coupling to the
EM current,
\begin{equation}
	\LL \supset -\frac{1}{4} F_{\mu\nu} F^{\mu\nu}
	- \frac{1}{4} F'_{\mu\nu} F'^{\mu\nu} + \frac{1}{2} m^2 A'_\mu
	A'^\mu
	- J^{\mu}_{\rm EM}(A_{\mu} + \kappa A'_{\mu})
\end{equation}
The interaction term $\LL \supset -\frac{1}{2}\kappa F_{\mu\nu}
F'^{\mu\nu} = - \kappa (\partial_\mu A_\nu) F'^{\mu\nu}$ is
equivalent, under integration by parts, to
$\kappa A_\nu \partial_\mu F'^{\mu\nu}$.
The Proca equation of motion for
$A'$ is $\partial_\mu F'^{\mu\nu} = m^2 A'^\nu + \OO(\kappa)$,
so to leading order in $\kappa$, we can write
the interaction term as 
$\LL \supset \kappa m^2 A_\nu A'^\nu$.

For a zero-velocity DM field, $A_0' = 0$, so the interaction
term is $- \kappa m^2 A \cdot A'$. This corresponds
to an effective current $J_{A'} = - \kappa m^2 A'$,
in analogy to the axion effective
current $J^{(a)} = g \dot a B$. The dark photon's velocity
dispersion means that the direction of $A'$
is not constant, but changes over the field's
coherence timescale, $\sim 10^6 m^{-1}$. Hence, dark photon DM
detection is analogous to axion DM detection,
with an `effective magnetic field' that varies
over the coherence timescale (and length scale) of the DM.

If we have a shielded experimental region, then
the sensitivity to $\kappa$ is set by that region's volume.
Unlike in the axion case, the frequency
of the forcing term in the interaction Hamiltonian is set by the
dark matter only, and does not depend on the target.
Correspondingly, the up-conversion and down-conversion scenarios do not
apply. The limit on the absorbed power is simply
\begin{equation}
	\bar P \lesssim \frac{\kappa^2 \rho V m^2}{\Delta m} \frac{\pi}{2}
\end{equation}
giving
\begin{equation}
	\bar N_{A'} \lesssim \kappa^2 \rho V t \frac{m}{\Delta m} \frac{\pi}{2}
\end{equation}
Note that $\bar N_{A'}$ does not vanish as $m \searrow 0$, unlike
most rates involving dark photons, since the DM field amplitude becomes
larger as $m$ becomes smaller (of course,
once $t \lesssim m^{-1}$, the expression becomes invalid). Putting in representative
numbers, 
\begin{equation}
	N_{A'} \lesssim \left(\frac{\kappa}{2 \times 10^{-19}}\right)^{-2} \frac{\rho}{0.3 \GeV \cm^{-3}} \frac{V}{\metre^3} \frac{t}{{\rm year}}
\end{equation}
for an order-1 mass range.
Consequently, for PQL or SQL-limited laboratory searches,
$\kappa \sim {\rm few} \times 10^{-19}$ is the smallest
kinetic mixing we could reasonably hope to detect.
If $m \lesssim L^{-1}$, and we are in the quasi-static
regime, then $\bar N_{A'} \lesssim \kappa^2 \rho_{A'} V t 
(m L)^2$.\footnote{This is true for e.g.\ a superconducting
shield. If electric fields are screened, but low-frequency magnetic fields can penetrate, as is the case
for a conducting cavity, then
the field inside the shield will depend on the response
of the material outside it to the DM field, potentially as
far away as the DM coherence length.
Experimental signatures of this effect
for very low-frequency dark photon DM
will be discussed in~\cite{peter}.}

Almost any axion experiment using a roughly-homogeneous, static background
magnetic field will act as a dark photon detector `for free', even
in the absence of the magnetic field. In particular,
cavity / dielectric haloscopes can again be $\OO(1)$ optimal
for dark photon conversion; existing cavity haloscopes
have set stringent limits on dark photon
DM in the $\sim \GHz$ frequency range~\cite{Arias:2012az}, while future proposals will often have sensitivity to dark
photon DM significantly before
they reach QCD axion sensitivity~\cite{TheMADMAXWorkingGroup:2016hpc,Baryakhtar:2018doz,Chaudhuri:2014dla}.
There are also many experiments and experimental proposals
for which the addition of a strong background magnetic
field would be practically difficult,
so they can detect dark photons but not axions
(e.g.\ experiments using superconducting 
phonon detectors, such as~\cite{Knapen:2017ekk,Griffin:2018bjn}).

In section~\ref{secStaticPower}, we noted how it is difficult
to obtain efficient conversion from dielectric
haloscopes at $\sim$ optical frequencies,
due to a lack of suitable low-loss,
high-$n$ dielectrics. These kinds of difficulties
are generic --- while achieving $\OO(1)$ of 
the $\bar P$ limit is fairly simple at microwave
frequencies, it is more challenging at
higher DM masses. In contrast, detecting
the converted excitations is often simpler,
due to the reduced blackbody noise and
more effective singe-quantum detectors.

In Figure~\ref{figdpsens}, 
we show some sensitivity limits and projections
for higher-mass dark photon DM detection
experiments, using different types of target excitations.
We compare these to the PQL-limited
sensitivities, for the appropriate target volumes and
integration times. These illustrate the conversion
efficiencies of the different targets and excitation
types (all of the schemes assume
efficient, almost background-free detection
of converted excitations).
This can help to identify how and where
improved targets can be found, versus where
larger volumes or longer integration times would
be necessary to improve sensitivity.

At $m_{A'} \gtrsim 10 \eV$, dark photon DM
would have enough energy to ionize atoms,
and such absorptions would be visible in WIMP
detection experiments. \cite{An:2014twa}
analysed the results of the Xenon10 experiment,
and used these to place limits on dark photon DM.
The dark photon absorption rate was calculated
using the imaginary part of the photon propagator
in liquid Xenon~\cite{An:2014twa}.
As can be seen from the PQL bound, LXe is quite an
inefficient dark photon absorber at
these frequencies --- the number of converted quanta
is a factor of $\sim 7000$ lower than an ideal
target occupying the same volume.

For slightly lower masses, dark photons
can have enough energy to excite
electrons in solid-state materials.
Depending on the material, this excitation
may be detected through ionisation
or scintillation signals~\cite{Bloch:2016sjj}.
In figure~\ref{figdpsens}, we show 
background-free, kg-year exposure projections
for sapphire and GaAs scintillators,
from~\cite{Griffin:2018bjn}. As the figure illustrates,
these materials are much more efficient at 
converting dark photon DM, with conversion
rates only a factor $\sim 5$ lower than ideal.

At lower frequencies, layered dielectrics have
been proposed as a way to convert
dark photon DM to photons~\cite{Baryakhtar:2018doz},
for detection using superconducting devices
such as TESs~\cite{Dreyling-Eschweiler:2014mxa,Dreyling-Eschweiler:2015pja,Cabrera1998,Karasik:2012rb,Lita:08,Bastidon:2015aha}, nanowires~\cite{Rosfjord:06} or MKIDs~\cite{Mazin,DayLeduc,GaoMazin}.
In figure~\ref{figdpsens}, we show a nominal
experimental projection using
eight different 100-period stacks,
each of area $\sim (18 \cm)^2$, covering a
decade in dark photon mass range
(each `chirped' stack can cover
a $\sim 30\%$ fractional mass range~\cite{Baryakhtar:2018doz}).
The material pairings used are
Si/${\rm Al}_2{\rm O}_3$, Si/Si${\rm O}_2$,
and Ti${\rm O}_2$/Si${\rm O}_2$.
These material choices give $\sim 1/30$ of 
the ideal conversion rate, corresponding to
the
$\frac{8}{\pi^2}\left(\frac{1}{n_2} - \frac{1}{n_1}\right)^2$ suppression from section~\ref{secStaticPower},
along with a $\sim 2/3$ misalignment factor
from the dark photon
polarization direction~\cite{Baryakhtar:2018doz}.

For even smaller dark photon masses, 
detectors with single-quantum sensitivity are
challenging, but it is possible that
superconducting technologies
such as TESs or nanowires could
achieve good enough energy resolution.
Given such detectors,
polar crystals have been proposed
as a target for dark photon conversion~\cite{Knapen:2017ekk,Griffin:2018bjn}.
Optical phonons in these crystals have gapped
dispersion relations, allowing non-relativistic
dark photons to convert to low-momentum optical
phonons without the need for further momentum-matching.
One drawback of this approach is that the
optical phonon dispersion relation is not easily
tuned, so absorption is concentrated at the
resonant frequency set by the optical phonon
dispersion relation (with quality
factor $\sim \OO(100)$~\cite{Griffin:2018bjn}).

In figure~\ref{figdpsens}, we show
the sensitivity projections from~\cite{Griffin:2018bjn}
for 1kg-year integrations with GaAs and sapphire crystals
(assuming efficient and background-free detection).
The resonant character of the absorption is clearly
visible. Averaging across the mass range covered, 
the GaAs crystal converts $\sim 250$ fewer photons
than an ideal target, while the sapphire crystal
is only a factor $\sim 25$ below optimal.



\section{Relativistic absorption}
\label{secRel}

For non-relativistic axion DM, we have seen that the
maximum absorption rate, averaged
over some axion mass range, scales with the magnetic
field volume (multiplied by the RMS magnetic field).
While coherent absorption can be helpful in terms of absorbing into
specific target modes (especially for background rejection
purposes), it does not result in parametrically enhanced rates compared
to incoherent absorption.
For example, the ionisation and electron-excitation
techniques discussed in the previous section have
the same scaling as dielectric haloscope absorption
(while the former were discussed in the context of dark 
photon DM detection, we could turn them into axion
experiments by placing them in a background magnetic field).
However, for absorption of relativistic axions,
coherent absorption can be parametrically advantageous.

In some situations, such as light-shining-through-wall
experiments~\cite{Sikivie:2007qm,Caspers:2009cj,Bahre:2013ywa,Betz:2013dza,Janish:2019dpr,Graham:2014sha}, axions
are produced at a precise, known frequency. In this case,
we want to concentrate the EM fluctuations
in the receiver into a small frequency range.
This can be accomplished using high-quality-factor
coherent modes, or with e.g.\ narrow linewidth
atomic transitions. Consequently, either `coherent' 
or `incoherent' absorption can be superior, depending 
on the linewidth achievable.

If the relativistic axion flux is spread over
an $\OO(1)$ range in frequencies, 
then the situation is different.
Decomposing the EM field in the experimental volume
into approximate plane wave modes,
each has overlap with a range of axion momenta
$\delta k \sim L^{-1}$. For axion masses
small enough, this corresponds to a
frequency range $\delta \omega \sim \delta k$,
even for axion masses $\OO(1)$ different from each other.
If we can concentrate the EM field
fluctuations for the appropriate mode into this frequency range,
this enhances them by $\sim \omega / \delta \omega \sim
\omega L$ over the broadband case, and so increases the absorption
rate.
In contrast, for a non-relativistic axion
axion DM signal, if we are $\OO(1)$-uncertain as to 
$m_a$, then for each spatial mode we want to absorb
into, the frequency uncertainty is $\OO(1)$
(we will assume that the background magnetic field is static).

For an axion with relativistic momentum $k \gg m_a$,
its frequency is
\begin{equation}
	\omega = \sqrt{k^2 + m_a^2} = k \left(1 + \frac{m_a^2}{2 k^2} + \dots\right)
\end{equation}
So, by varying the axion mass by $\OO(1)$,
we vary $\omega$ by $\sim m_a^2 / (2 \omega^2)$.
Consequently, we need $m_a^2 \lesssim \omega/L$
to realise the full $\sim \omega L$ enhancement from above.

This enhancement for relativistic absorption is
actually very simple to implement. If we know
the approximate direction that the relativistic
axions will be coming from, then we can construct
a long tube facing in that direction, and fill it
with an approximately uniform transverse magnetic
field. This is exactly the setup of the CAST
experiment~\cite{Anastassopoulos:2017ftl}, which
looks for axions produced inside the Sun.
If the axion mass is small enough such that
$k_a - k_\gamma = k_a - \omega < L^{-1}$, i.e.
$m_a^2 / (2 \omega) \lesssim L^{-1}$, then axions
travelling down the tube convert coherently to
photons, with conversion probability $\propto (g B
L)^2$, where $L$ is the length of the tube. Thus, if $F_a \sim a^2 \omega$ is the axion
flux, and $A$ is the tube's cross-sectional area, then the rate of converted photons is
\begin{equation}
	\Gamma_\gamma \sim F_a A (g B)^2 L^2 \sim (g a B)^2 (A L) (\omega L)
\end{equation}
Comparing this to equation~\ref{eqpbar1}, for a static-field experiment absorbing
non-relativistic axions over an $\OO(1)$ mass range, we see that the relativistic case is enhanced
by a factor $\sim \omega L$. This is just the number of wavelengths over
which we can build up coherently, as expected. Photons
in a mode with a specific momentum naturally have a
specific frequency, realising the concentration
of the fluctuation spectrum that we wanted.

If $k_a - k_\gamma > L^{-1}$, then we would need to modify
the photon dispersion relation inside the tube
to obtain coherent conversion.
For example, as done with CAST~\cite{Arik:2008mq,Arik:2015cjv,Arik:2011rx,Arik:2013nya}, we could introduce some
gas into the tube, changing the refractive index for (X-ray) photons.
In this case, the maximum possible enhancement, if $m_a$
is $\OO(1)$ uncertain, is $\sim \omega^2 / m_a^2$.
For $\omega \sim m_a$, the enhancement disappears,
as expected from our analysis of non-relativistic DM.

If we allow the background magnetic field to oscillate in time,
we could improve the theoretical detectability further
by using the down-conversion ideas discussed above.
However, the largest naturally-occurring flux of relativistic
axions is from the Sun, and is dominantly
at $\sim \keV$ energies~\cite{Andriamonje:2007ew}, so creating a suitable background field
is not technologically plausible.

\subsection{Dark photons}
\label{secdprel}

\begin{figure*}[t]
	\begin{center}
		\includegraphics[width=.7\linewidth]{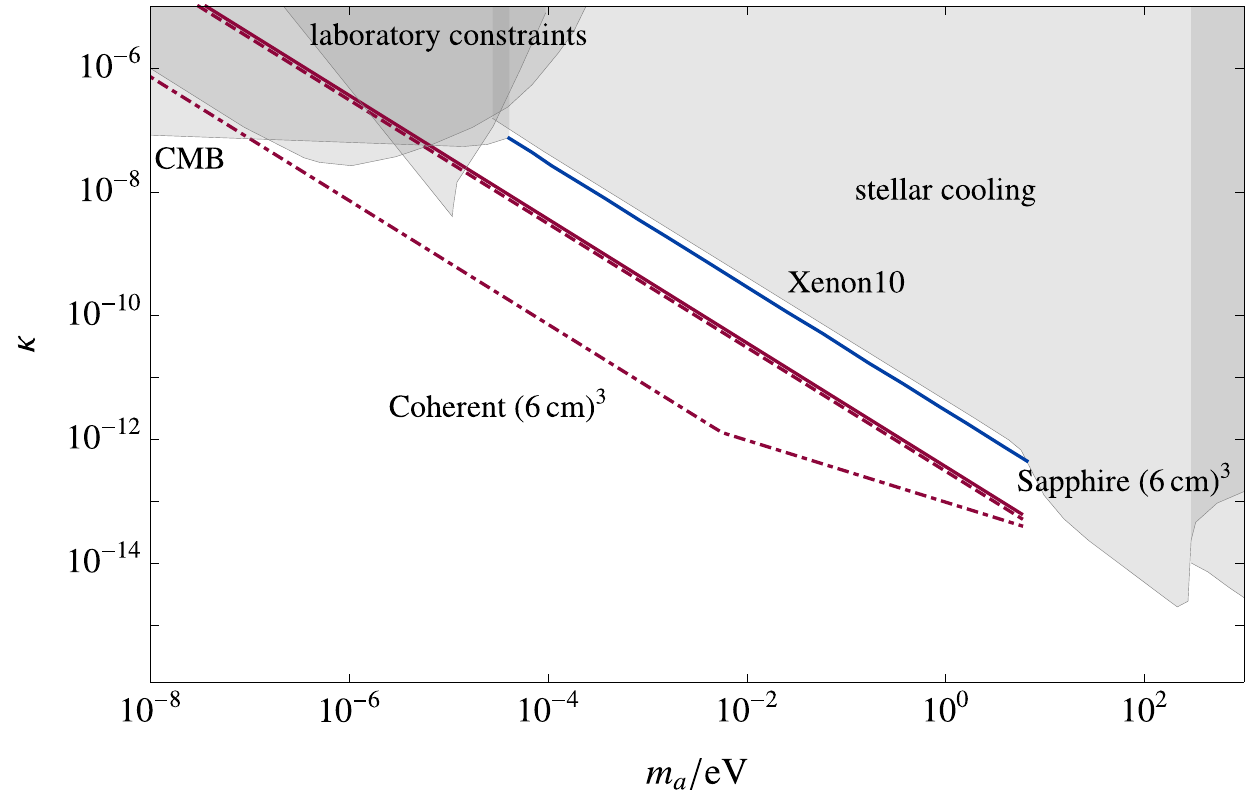}
		\caption{Sensitivity of dark photon absorption
		experiments to the flux of longitudinal dark
		photons from the Sun. At $m \ll 300 \eV$, this
		flux is dominated by longitudinal emission~\cite{Redondo:2013lna}, so we only consider absorption of longitudinal dark photons.
		The `Xenon10' line shows the limits
		from dark photon absorption in the Xenon10
		experiment~\cite{An:2013yua}.
		The solid red line shows the sensitivity 
		for a sapphire target, with 1kg-year exposure
		(as in figure~\ref{figdpsens}). The dashed red line shows
		the constraints from an optimal incoherent absorber
		with the same volume ($\sim (6 \cm)^3$),
		illustrating that sapphire is an almost optimal absorber, and in particular, much more efficient
		than liquid xenon (Xenon10 uses
		a $\sim (17 \cm)^3$ volume).
		The dot-dashed line shows the maximum theoretical
		absorption rate for a cube of the same volume,
		taking advantage of coherent enhancement.
		As discussed in section~\ref{secdprel}, realising this
		enhancement in a practical experiment would
		likely be very difficult.
		The gray regions correspond to existing constraints
		on dark photons from other observations~\cite{An:2013yfc,Vinyoles:2015aba,An:2013yua,Betz:2013dza,Mirizzi:2009iz,Jaeckel:2008fi,Bartlett:1988yy,Williams:1971ms}.
		} 
	\label{figdpsun}
	\end{center}
\end{figure*}

The case of relativistic dark photons is made more complicated
by the different behaviour of longitudinal and transverse
excitations (for the non-relativistic case,
this difference is less important).
For a plane wave travelling in the $z$ direction,
the transverse polarization vectors for $A'_\mu$ are spanned
by $\epsilon_\mu^{(x)} = (0, \hat x)$ and
$\epsilon_\mu^{(y)} = (0, \hat y)$,
while the longitudinal polarization vector
is $\epsilon_\mu^{(L)} = (-|k|,\omega \hat z)/m$
(for an on-shell excitation)~\cite{Hardy:2016kme}.
For a vector coupling to a generic, non-conserved SM
current, the $\sim \omega/m$ enhancement of
the longitudinal polarization vector for relativistic excitations
leads to energy-enhanced longitudinal emission~\cite{Dror:2017nsg}. 
Of course, the EM current that the dark photon couples
to is conserved, so we instead expect the longitudinal mode
to decouple as
$m \rightarrow 0$;
the $m^2$ factor in the coupling term $\LL \supset \kappa
m^2 A^\mu A'_\mu$ ensures that rates still decrease
at least as fast as $m^2$ for $m \rightarrow 0$.

The maximum rate for the absorption
of transverse $A'$ is
\begin{equation}
	\Gamma \lesssim \begin{cases}
		\kappa^2 \frac{m^4}{\omega^4} F V \omega (\omega L) &
		\mbox{for } m^2 \lesssim \omega/L \\
		\kappa^2 \frac{m^2}{\omega^2} F V \omega &
		\mbox{for } \omega/L \lesssim m^2 \lesssim \omega^2 
	\end{cases}
\end{equation}
where $F$ is the dark photon flux,
in analogy to the axion analysis above.
Since SM photons are transversely polarized, it is simple 
to obtain the $(\omega L)$ and $\omega^2/m^2$ enhancements
corresponding to these cases.
For $m^2 \lesssim \omega/L$, the momentum difference
between $A'$ and free-space photons is small
compared to the inverse size of the experiment,
so a conducting cavity achieves this conversion rate
(cf.\ the dark photon bounds from CAST~\cite{Redondo:2008aa}).
To obtain a momentum match for $m^2 \gtrsim \omega/L$,
we would need to modify the dispersion relation of the
photon inside the volume, e.g.\ using a gaseous
or liquid medium,
similarly to CAST~\cite{Arik:2008mq,Arik:2015cjv,Arik:2011rx,Arik:2013nya}.

The maximum absorption rate for
relativistic longitudinal dark photons
is $\sim \omega^2/m^2$ times larger
than the transverse rates above.
However, to convert longitudinal $A'$ to 
transversely polarized SM photons, we would need
an anisotropic medium to set the photon
polarisation direction. If the medium is dense
enough to accomplish this conversion efficiently,
it will generally modify the dispersion relation
of the SM photon as well. This would spoil the momentum
match between the dark photon and the SM
photon, and we would no longer get coherent conversion.
While there are theoretically ways to get around this
--- e.g.\ by multiplexing many wavelength-sized cavities
together with the appropriate phase offsets ---
these seem difficult to implement at large enough $\omega$,
so that the $\omega L$ coherence enhancement is significant
for a laboratory-scale experiment.

If we do not take advantage of coherent conversion,
then the maximum absorption rate
for relativistic longitudinal dark photons
is 
\begin{equation}
	\Gamma \sim \kappa^2 \frac{m^2}{\omega^2} F V \omega
\end{equation}
This is $\sim (\omega/L)/m^2$ times larger than
the coherence-enhanced transverse rate for $m^2 \lesssim \omega/L$,
and equal to the transverse rate for $\omega/L \lesssim m^2 \lesssim \omega^2$. Consequently, even without coherence enhancement,
the theoretical limit on the absorption rate 
 of longitudinal dark photons is always as
fast or faster than the coherence-enhanced absorption rate
of an equivalent flux of transverse dark photons.

For $m$ much less
than the plasma frequency in Sun ($\omega_p \sim 300 \eV$
in the solar core), the production rate of longitudinal
$A'$ from the Sun is $\propto m^2$, whereas
the rate for transverse $A'$ is $\propto m^4$.
Consequently, for $m \ll 300 \eV$, the solar
flux is dominated by longitudinal $A'$~\cite{Redondo:2013lna}.
This means that longitudinal absorption is even more favoured.
Limits on this dark photon flux have been set 
by looking for longitudinal mode absorption
in the Xenon10 detector~\cite{An:2013yua}. However,
the $A'_L$ flux from the Sun peaks
at $\omega \sim 10 - 100 \eV$, and
as we saw in section~\ref{secDP}, liquid xenon
is a rather inefficient absorber
of dark photons at these frequencies.
If we used a more efficient absorption process,
such as electronic excitations in sapphire,
then significantly stronger constraints could
be set. 
Figure~\ref{figdpsun} shows the projected constraints
from dark photons from the Sun, using the same
1kg-year sapphire experiment considered for DM detection
in figure~\ref{figdpsens}.
The dashed line shows the theoretical sensitivity limit
for incoherent absorption, using the same volume
and integration time, illustrating that the sapphire
is a close-to-optimal absorber.
The dot-dashed line shows the additional sensitivity
we could theoretically gain by taking 
advantage of coherent absorption
of longitudinal $A'$ (though we do not have
a practical experimental proposal for this).

As well as the Solar flux, there are other potential sources
of relativistic dark photons, including dark radiation
produced in the early universe.
If this has a temperature comparable to the CMB, then
most of its power is at $\sim$ mm wavelengths,
so there would be the potential for $\omega L \sim 10^3$
coherence enhancements in the converted power,
for a meter-scale experiment. The polarization
of the dark radiation would depend on the early universe
production mechanism, and would need to be considered
in designing an experiment.


\section{Conclusions}

In this paper, we have derived limits on the experimental detectability
of axion DM, assuming that it couples to the SM through
the $a F_{\mu\nu} \tilde F^{\mu\nu}$ operator. Similar
analyses can be applied to other DM-SM couplings; most simply,
to kinetically mixed dark photons, as we considered 
in section~\ref{secDP}.
For DM candidates which do not couple directly
to the EM field, the EM sum rule based bounds
we derived do not apply as straightforwardly. However,
we can still use fluctuation-based
(and/or energy absorption) analyses 
to bound the sensitivity of experimental searches. 

In some cases,
these can be related to the EM field fluctuations.
For example, if DM consists of a vector which
couples to the SM $B-L$ current, then its couplings
to electrons and protons are the same as those
of a dark photon with 
$\kappa_{\rm eff} = g_{B-L} / e$ (though its
coupling to neutrons is different).
So, in circumstances where neutrons are not important
(for example, in dielectric haloscope detection schemes,
where electrons dominate the dielectric response),
the dark photon limits we derived will apply.

Analysing the $B-L$ coupling directly,
the PQL sensitivity is set by the fluctuations of the
$B-L$ current in the target system. 
Since,
for a shielded volume, the EM current fluctuations
are related to the EM field fluctuations,
and -- when neutrons are unimportant -- 
the $B-L$ current is proportional to the EM current,
the fluctuations of the $B-L$ current can
be related to those of the EM field.
For example, if $e$ were smaller, than the fluctuations
of the electron-number current in the shielding would have 
to be larger in order to cancel out the 
the EM field fluctuations, corresponding to the 
$\kappa_{\rm eff} \propto 1/e$ behaviour.

Of course, since there is not a perfect degeneracy
between the effects of a $B-L$ vector and EM fields,
the dark photon type limits can be circumvented.
In particular, one can look for the force
exerted on neutron-containing objects, which can
give good sensitivity at small DM masses~\cite{Graham_2016}.

A $B-L$ vector is an example of how
there can be links between the EM field fluctuations, and
the fluctuations of different operators.
Such links may also be present in other cases,
e.g.\ axion-like DM coupling to fermions,
which for slow-moving fermions acts
like an effective magnetic field. We leave
investigations of these and other examples to future work.

As discussed in Section~\ref{secRel}, similar
analyses to those in this paper can be applied
beyond the scenario of DM searches, to other kinds
of weakly-coupled particle detection.
In particular, quantum measurement theory has
been extensively developed by
the gravitational wave detection community (see e.g.~\cite{Miao}).
Similar analyses can also be applied to DM scattering
(rather than absorption) experiments, and to experiments
looking for the production of new states.

Returning to the case of axion DM detection,
our analyses clarify the limits on low-frequency axion
detection, and motivate approaches for getting
around the quasi-static suppression.
In a companion
paper~\cite{srf}, we perform a basic analysis of microwave
up-conversion experiments in this regime, 
considering their sensitivity in light of plausible practical
limitations. 
Alternatively, one could try to enhance the sensitivity
of static-background-field searches,
via active or passive approaches.
We leave investigations of
such setups to further work.
More generally, it would also be interesting to understand
whether violating the $\omega \tchi_i (\omega) \ge 0$
assumption, e.g.\ using unstable systems~\cite{Miao:2015pna},
could help improve sensitivity.


\begin{acknowledgments}
 We thank Masha Baryakhtar, Saptarshi Chaudhuri, Peter Graham, Junwu
Huang, Kent Irwin, and Jesse Thaler for useful discussions.
This research was supported in part
by the National Science Foundation under Grant No. PHYS-1720397, the Gordon and Betty Moore Foundation Grant GBMF7946, and
	by the Munich Institute for Astro- and Particle Physics (MIAPP) which is funded by the Deutsche Forschungsgemeinschaft (DFG, German Research Foundation) under Germany's Excellence Strategy – EXC-2094 – 390783311.

\end{acknowledgments}


\appendix

\section{Quantum measurement theory}
\label{appQmeas}

Almost all production mechanisms for light ($m \lesssim$ few eV)
bosonic dark matter result in its state today being
a coherent, classical-like oscillation, with high occupation
number --- for examples, see e.g.~\cite{Kawasaki:2014sqa,Ringwald:2015dsf,Co:2017mop,diCortona:2015ldu,Graham:2015rva}.
Since its couplings to a SM laboratory system are generally constrained
to be very weak --- in particular, far too weak for the interaction
to significantly affect the DM state --- it is valid to treat
the DM as a fixed classical background field, for the purposes of detection.
The question then becomes how well we can detect a small,
classical forcing adding to the SM Hamiltonian.
The ultimate limits on doing this are set by quantum mechanics, 
and are one of the subjects of quantum measurement theory~\cite{wiseman_milburn_2009}.
Here, we give a brief overview of some basic results that we will
use.

Setting up the problem,
suppose that we have a quantum system, interacting with a classical oscillation
$j(t)$ through the interaction Hamiltonian $H_{\rm int} = g j(t) \hat X$,
where $\hat X$ is some operator on the system, and $g$ is a coupling that
we will take to be small. If we treat our quantum system in the interaction
picture, and start with it in a state $|\psi \rangle$ at time
$t_0$, then at leading order in $g$,
by $t_1$ it has evolved to
\begin{align}
	|\psi_I(t_1)\rangle &\simeq \left(1 - i g \int_{t_0}^{t_1} dt \, j(t) \hat X(t) \right) |\psi\rangle
	\equiv \left(1 - i g \hat V\right) |\psi\rangle
\end{align}
The overlap of $|\psi(t_1)\rangle$ with $|\psi\rangle$ is therefore
\begin{equation}
	|\langle \psi(t_1)|\psi\rangle| \simeq 1 - \frac{1}{2}g^2 \langle \psi | 
	(\Delta \hat V)^2 |\psi \rangle
	\label{eqQCR}
\end{equation}
where $\Delta \hat V \equiv \hat V - \langle \psi | \hat V | \psi
\rangle$. Consequently, the ability to distinguish between
the presence and absence of a weak signal is set by the fluctuations
of the interaction Hamiltonian~\cite{Braginsky:1999zr}.
This is related to the theory of `Quantum Cramer-Rao bounds'~\cite{Tsang:2011zz,Miao:2017vmx,Pang:2019ztr},
and is sometimes referred to as an `Energetic Quantum
Limit' or `Fundamental Quantum Limit' (FQL).

In our cases, the DM signal $j(t)$ will generally be narrow-bandwidth,
and it will be more useful to go to a spectral representation.
Taking the same assumptions as above,
\begin{equation}
	|\langle \psi(t_1) | \psi \rangle|
	\simeq 1 - g^2 \int_0^\infty d\omega \, |\tilde j_t(\omega)|^2
	\SC_{\Delta X \Delta X}(\omega)
\end{equation}
where $j_t(t) = j(t) \mathbf{1}_{t_0 < t < t_1}$,
and $\SC_{\Delta X \Delta X}$ is the symmetrised
spectral density of $\Delta X$.
In the simplest case, where $j(t)$ is a single-frequency oscillation,
$j(t) = j_0 \cos (\omega t)$, we have
\begin{equation}
	\PPex \simeq \frac{1}{2} g^2 j_0^2 \SC_{\Delta X \Delta X}(\omega) t_{\rm exp}
\end{equation}
where $t_{\rm exp} = t_1 - t_0$,
and $\PPex \equiv 1 - |\langle \psi(t_1)|\psi\rangle|^2$ is the probability of changing the detector system's state.
More generally, if we can treat $j(t)$ as a stochastic process,
with some power spectral density $S_{jj}$, then in the limit
where we evolve for a time long compared to the inverse bandwidth of spectral features in
$S_{jj}$, 
\begin{equation}
	\langle \PPex \rangle \simeq \frac{g^2 t_{\rm exp}}{\pi} \intoinf d\omega \, S_{jj}(\omega) \SC_{\Delta X \Delta X}(\omega)
\end{equation}

To gain some intuition for these results, it is helpful to consider
the case where our system is a harmonic oscillator, coupled
to an external force, $\hat X = \hat x$. We will assume that the oscillator
has some small coupling to other degrees of freedom,
giving it a high quality factor $Q \gg 1$.
In the ground state, we have
\begin{equation}
	\SC_{xx}(\omega) \simeq \frac{Q}{1 + Q^2  \frac{(\omega^2 - \omega_0^2)^2}{ \omega_0^4}} \frac{1}{\omega_0} \frac{1}{M \omega_0}
\end{equation}
for $\omega$ close to $\omega_0$, where $\omega_0$ is the natural 
frequency of the oscillator, and $M$ is its mass.
Here, $\frac{1}{M \omega_0}$ is the squared position uncertainty.
By e.g.\ decreasing $M$ while keeping $\omega_0$ fixed, we increase
the position uncertainty, and so decrease the momentum uncertainty. 
Since the external force changes the momentum of the system,
having smaller momentum fluctuations helps to detect the forcing.

This illustrates how, for Gaussian states, the dependence on the fluctuations
of $\hat X$ has a simple interpretation in terms of conjugate variables.
However, things do not have to be that simple.
For example, if the harmonic oscillator is initially in a number
state, then
$\langle n | \hat x^2 | n \rangle = \frac{2n+1}{2 M \omega_0}$, and
$\langle n | \hat p^2 | n \rangle = \frac{M \omega_0}{2} (2n+1)$ --- both the position
and the momentum uncertainties are higher than for the ground state.
However, by e.g.\ measuring the energy of the final state,
we can still attain
the $\SC_{xx}$ bound~\cite{Caves:1980rv}.
Effectively, the rate for absorption (and emission)
of quanta due to the forcing is Bose-enhanced by the initial occupation number.
This provides an example of how, even for more complicated systems, 
the $\SC_{\Delta X \Delta X}$ prescription still gives the correct answers.

In the discussion above, we assumed that the detector system
was allowed to evolve for a time $t_1 - t_0$, and only measured at the
end. For many experimental setups, something
closer to continuous monitoring is implemented --- e.g.\
in resonant cavity experiments, the output port is connected
to an amplifier. However, since we are only
concerned with the fluctuations of $\Delta X$, 
as long as we include the rest of the system's dynamics
in determining these (including measurements,
feedback etc), this does not present a problem
(cf the discussion of deferred measurement in~\cite{Tsang:2011zz}).

The above limits were based on knowing precisely which quantum state
$|\psi\rangle$ our system starts in, and precisely how it would evolve
from there.
Other `fluctuations', due to our uncertainty about the system's state
(e.g.\ thermal fluctuations) have the opposite effect, making it harder
to tell whether a signal is present. 
In some circumstances, $j(t)$ itself may be uncertain --- for example,
the Fourier components for a virialized DM signal
are expected to have random amplitudes and phases.
If these unknown Fourier amplitudes affect the system's response to
the signal, then the effective SNRs for the
independent components generally add in quadrature~\cite{Helstrom1},
rather than linearly, as per the Dicke radiometer formula~\cite{Dicke:1946glx}.

\subsection{Linear amplifiers}
\label{seclinamp}

The FQL detectability limit discussed above applies
to any type of detection system, as long as we properly
calculate the quantum fluctuations of $\hat X$.
However, in many cases, sensors are complicated non-equilibrium
devices, and the fluctuations they cause may be difficult
to compute. In addition, detection schemes may fail
to obtain the FQL.
Consequently, it is often helpful to
consider the sensitivity limits for more restricted classes
of sensors.

A common example of such a sensor, relevant to many axion
detection experiments, is a linear amplifier.
There is an extensive literature on the quantum
theory of linear amplifiers (see e.g.~\cite{Clerk_2010} for
a review). In
many circumstances, it is a good approximation
to treat the `fluctuations', both quantum
and statistical, of measured and output
quantities as Gaussian. Then, the relevant quantities
can be summarised as `noise' spectral densities.

We will denote the PSD of backaction
noise acting on $\hat X$ as $\bar S_{FF}(\omega)$,
and the output imprecision noise (referred
back to $X$) as $\bar S^I_{XX}(\omega)$.
The amplifier does not have to be connected `directly'
to the $X$ degree of freedom for this description
to make sense, so long as the whole system behaves
linearly. A common setup is where a high-power-gain
amplifier is coupled weakly to the target system
(where `weakly' means that it has a very subdominant
effect on the system's damping). This is referred to
as `op-amp' mode~\cite{Clerk_2010}. In this scenario,
the added noise associated with
the amplification process is~\cite{Clerk_2010}
\begin{equation}
	\bar S^{\rm add}_{XX}(\omega) = |\tilde \chi (\omega)|^2 \bar S_{FF}(\omega)
	+ \bar S^I_{XX}(\omega) + 2 {\rm Re} [\tchi(\omega) \bar S^I_{XF}(\omega)]
	\label{eqsxxadd}
\end{equation}
where $\bar S^I_{XF}$ denotes the correlation 
of the backaction noise, and the output imprecision noise
referred back to $X$.
The total output `noise', referred back
to $X$, is
	$\bar S^{\rm tot}_{XX} = \bar S^{\rm add}_{XX} + \bar S^{n}_{XX} + \bar S^{\rm ZPF}_{XX}$,
where $\bar S^{\rm ZPF}_{XX}(\omega) = \tchi_i(\omega)$ corresponds to the zero-point
fluctuations of $X$, and $\bar S^n_{XX}$ summarises
the other noise contributions (e.g.\ for thermal
noise, $\bar S^n_{XX}(\omega) = 2 n_T(\omega) \tchi_i$).
If we are attempting to detect a signal
$j$, whose Fourier components have unknown
phases, and integrate for a time long compared to
inverse spectral bandwidths, then the expected
${\rm SNR}$ squared is 
\begin{equation}
	{\rm SNR}^2 =  g^4 t \intoinf \frac{d\omega}{2\pi} \left(\frac{\bar S_{jj} |\tilde \chi|^2}{\bar S^{\rm tot}_{XX}}\right)^2
\end{equation}
To maximise our SNR, we want to reduce $\bar S^{\rm add}_{XX}$.
For a high-gain amplifier, this is bounded
below by 
$\bar S^{\rm add}_{XX}(\omega) \ge |\tchi_i(\omega)|$~\cite{Clerk_2010}.
Achieving this `quantum limit' requires
$\bar S_{IF}(\omega) = -\frac{\tchi_r(\omega)}{2 \tchi_i(\omega)}$,
i.e.\ that the correlations between
the imprecision and backaction noise are
set by the phase of the response function.

In many circumstances, it is easier to implement
linear amplifiers with \emph{uncorrelated} 
imprecision and back-action noise. Following
the gravitational wave detection literature~\cite{Pang:2019ztr},
and papers such as~\cite{Kampel:2017gze}, 
we will refer to this as the `Standard Quantum Limit'
(SQL), as opposed to the `quantum limit' (QL) in
which correlations are permitted.
Generally, it is the case that 
\begin{equation}
	\bar S_{XX}^I \bar S_{FF} - |\bar S^I_{X F}|^2
	\ge \frac{1 + \Delta
	\left[2 \bar S^I_{X F} \right]}{4} 
\end{equation}
where $\Delta[z] \equiv (|1 + z^2| - (1 + |z|^2))/2$~\cite{Clerk_2010}. This implies that $\bar S^I_{XX} \bar S_{FF} \ge \frac{1}{4}$.
If there are no correlations, then
$\bar S_{XX}^{\rm add} \ge |\tchi|^2 \bar S_{FF} + \frac{1}{4 \bar S_{FF}}$,
which is minimised by $\bar S_{FF} = \frac{1}{2 |\tchi|}$,
giving $\bar S_{XX}^{\rm add}(\omega) \ge |\tchi(\omega)|$.

These limits apply to detection schemes which are
invariant under time translation, usually referred
to as `phase-invariant' (i.e.\ they treat sine
and cosine signals in the same way). By varying e.g.\ the
detector coupling in a time-dependent way, 
sensitivity-enhancing schemes such as backaction
evasion can be implemented~\cite{Clerk_2010}.

The SNR limits look rather different to the
fluctuation-based FQL limits discussed above.
However, for equilibrium targets in the linear
response regime, $\tchi(\omega)$ is related
to $\bar S_{\Delta X \Delta X}(\omega)$ via the 
fluctuation-dissipation relations.
As we will
see in the text, this can lead to closely-related
FQL and SNR bounds.

In some circumstances, we will want to go beyond the 
op-amp regime, and couple the detector more strongly to the
target system. We discuss this in~\ref{secIsolated}.



\section{Atomic and molecular magnetic fields}
\label{apmag}

In the main text, the `background' magnetic field $B_0$ was generally
taken to be a smoothed version, not taking into account
the large magnetic fields inside molecules, atoms, nuclei,
etc. This is justified since, if the electric
field associated with signal excitations is slowly-varying in space,
then the interaction strength only depends on the magnetic multipole
moments of sub-wavelength structures. For example, if we integrate
over a volume containing some currents and spins,
and the magnetic fields from sources outside
the volume are small, then
\begin{equation}
	\int dV B = \frac{2}{3}m_{\rm tot}
\end{equation}
where $m_{\rm tot}$ is the total magnetic dipole moment of the
matter~\cite{Jackson}. 

For the $\lesssim \eV$ excitations we considered,
the smallest scale of spatial variation is e.g.\
motions of atoms in a lattice, or molecular vibrations.
In particular, these are above the atomic scale. So,
the magnetic field strength is, at best, that arising from
atomic magnetic dipoles, which give a $\sim$ Tesla field.
Thus, we cannot gain an advantage over using a
strong, roughly uniform background magnetic field,
which can be of multi-Tesla strength.



\bibliography{axionCavity}

\end{document}